\documentclass[prd,aps,showpacs,nofootinbib,floats,floatfix,twocolumn,letterpaper]{revtex4}

\usepackage{amssymb,graphicx}
\usepackage{epsfig}
\usepackage[usenames]{color}

\def\apjl{Ap. J. Letters}
\def\mnras{MNRAS}

\newcommand{\had}{{\sc had}}

\begin{document}

\title{Perturbed disks get shocked. Binary black hole merger effects on accretion disks}

\author{Miguel Megevand${}^1$,  Matthew Anderson${}^2$,
  Juhan Frank${}^1$, Eric W. Hirschmann${}^3$,  Luis Lehner${}^1$,
Steven L. Liebling${}^4$, Patrick M. Motl${}^5$ and David Neilsen${}^3$}

\affiliation{
${}^1$Department of Physics and Astronomy, Louisiana State
University, Baton Rouge, LA 70803-4001, \\
${}^2$Department of Mathematics, Brigham Young
University, Provo, UT 84602,\\
${}^3$Department of Physics and Astronomy, Brigham Young
University, Provo, UT 84602,\\
${}^4$Department of Physics, Long Island University--C.W. Post Campus,
Brookville, NY 11548, \\
${}^5$Department of Natural, Information and Mathematical Sciences,
Indiana University Kokomo, Kokomo, IN 46904
}

\date{\today}

%
%
\begin{abstract}
The merger process of a binary black hole system can have a
strong impact on a circumbinary disk. In the present work
we study the effect of both central mass reduction (due to the
energy loss through gravitational waves) and a possible black hole
recoil (due to  asymmetric emission of gravitational radiation).
For the mass reduction case and recoil directed along the
disk's angular momentum, oscillations are induced in the disk
which then modulate the internal energy and bremsstrahlung luminosities.
On the other hand, when the recoil direction has a component orthogonal
to the disk's angular momentum, the disk's dynamics are strongly impacted,
giving rise to relativistic shocks.  The shock heating leaves its signature in
our proxies for radiation, the total internal energy and bremsstrahlung luminosity.
Interestingly, for cases where the kick velocity is below the smallest orbital
velocity in the disk (a likely scenario in real AGN),
we observe a common, characteristic pattern in the internal energy of the disk.
Variations in kick velocity simply provide a phase offset in the characteristic
pattern implying that observations of such a signature could yield
a measure of the kick velocity through electromagnetic signals alone.
\end{abstract}

\maketitle

%
%
\section{Introduction:}
The study of a number of astrophysical systems will soon add
gravitational wave astronomy as a new tool to complement observations
in the electromagnetic band. Since most systems capable of producing detectable
gravitational waves will also radiate strongly in the electromagnetic
band (see, e.g., \cite{Sylvestre:2003vc,Stubbs:2007mk}), combining information
from both spectra 
will allow for a richer description of these systems. Furthermore, the complementary
nature of observation in both bands will help the detection enterprise as
a signal in one band will help follow up studies in the
other (see, for instance, ~\cite{Haiman:2008zy,Bloom:2009vx}.)

Among interesting possible sources of strong signals in both spectra, the collision
of a binary black hole system within a circumbinary disk presents the possibility of
a detection of gravitational waves (as the black holes merge), which will be followed by
electromagnetic signals emitted by the disk as its dynamics are affected in the process~\cite{Milosavljevic:2004cg}.
This scenario is common in nature, since massive black holes exist in the core of
most galaxies and galaxies undergo mergers throughout their evolutionary path.
As galaxies merge, they produce a binary black hole in the newly formed
galaxy which eventually collide as their orbit shrinks through several mechanisms.
As discussed in \cite{Milosavljevic:2004cg}, a circumbinary disk is formed
as the binary hollows out the surrounding gas, and the disk becomes mostly
disconnected from the binary's dynamics~\cite{2002ApJ...567L...9A,2003MNRAS.340..411L}.
Afterwards, while the disk remains essentially frozen, the black holes' orbits continue to shrink until
they merge.

The merger process gives rise, in particular, to two relevant effects that
will perturb the disk (see, e.g.,~\cite{Milosavljevic:2004cg,Kocsis:2008va}). One is
related to the final mass of the black hole, which is less than the initial total mass as the system radiates
energy via gravitational waves\footnote{Possible observable consequences of this effect
were first discussed in~\cite{bodephinney}.}. The other one is a consequence of the radiation of linear momentum, which if
asymmetric (as in the case of an unequal mass binary, or asymmetric individual angular momenta of the black holes),
induces a nontrivial recoil on the nascent black hole.
This recoil effect has been predicted before through perturbative analysis of
Einstein equations \cite{kickold1,kickold2}, and recent numerical simulations implementing
the equations in full show even higher recoil velocities are possible
\cite{Koppitz,kick1,kick2,kick3,kick4,kick5,kick6,kick7,kick8,kick9}.
The largest recoil velocities found correspond to mass ratios close to 1
and spins lying anti-aligned on the orbital plane. In the case of quasicircular orbits,
recoil velocities up to about 4000~km/s have been calculated
\cite{kick3}. However, most of the black
hole collisions occurring in nature are expected to produce kicks of
about 500~km/s or less, since larger kicks would occur only in the case of
nearly equal masses \cite{distributions}.

As a result of both effects mentioned above, the fluid dynamics in the disk is modified and shocks
may be induced. The shocks' energy can then heat the gas, which can produce electromagnetic flares.
These flares are expected to occur later (a few months to years), and to last
considerably longer (thousands to hundreds of thousands of
years)\cite{kickdisk1,kickdisk2,kickdisk3}, suggesting
tantalizing prospects for LISA observations aiding and complementing the
electromagnetic observational prospects of these systems.
For the recoiling black hole case, prior studies, which employ simulations of
collisionless particles in Keplerian orbits forming a flat (zero height) disk,
predict emissions ranging from UV to x-rays~\cite{kickdisk2,kickdisk1} or in
the infrared~\cite{kickdisk3} if this radiation is assumed to be absorbed
before leaving the disk and re-emitted.
Since these studies employ a particle description of the fluid, they
can not fully capture the development (and hence influence) of shocks, which
must be estimated by detecting collisions between particles. A recent work \cite{massloss}
adopted a field description for the fluid and studied the impact of a mass reduction
in a pseudo-Newtonian potential to account for an innermost stable circular
orbit (ISCO) at $r=6M$ (which corresponds to the ISCO of
a nonspinning black hole, while this is a rather
uncommon output~\cite{Buonanno:2007sv,Boyle:2007ru,Rezzolla:2007rz,Lousto:2009mf} in the merger of two black holes,
the spin value will play a relevant role mainly if accretion develops).  Based on computations of bremsstrahlung luminosity, that work predicts
a decrease in luminosity as the fluid orbits adjust to the reduced gravitational
potential.

In this work we study the effects on the disk by also considering
a perfect fluid but in our case we do so employing the fully relativistic hydrodynamic
equations in a background space-time. Thus, we are able to examine effects of spin, mass
reduction and accretion, and comment on the relevance of different processes.
In particular, our studies indicate that a significant distortion of the disk develops
as time progresses when the kick has a component perpendicular to the disk's axis and that
qualitatively similar features are present in all these cases.

In Section II,
we briefly review our formulation of the problem and numerical approach.
Section III describes our initial configuration. We discuss the observed
dynamics in the disk after the merger has taken place in section IV, taking
into consideration the effect of mass reduction and different recoil velocities.
Section V concludes and offers some further considerations.

Unless otherwise specified, we use geometrized units, where $G=c=1$, and sum
over repeated indexes.
Greek-letter indices range from 0 to 3.

%
%

\section{Overview of the numerical approach:}
We implement the general relativistic MHD equations using a high resolution
shock capturing module described in \cite{had1,had2}. We introduce however
a slight modification of
the hydrodynamic equations inside the horizon to improve the fluid's behavior close
to the excision region. 
Given an equation of the form\footnote{For clarity we use here a simple
  expression to represent the fluid equations. See \cite{had1,had2} for the
  full equations.}
\begin{equation}
\dot U + F(U)' =  S,
\end{equation}
 we modify it in order to include a damping term:
\begin{equation}
\dot U + F(U)' =S -f(r) {(\Delta x)}^p (U-U_0),
\end{equation}
 where the function $f(r)$ decreases smoothly with $r$,
 from $100$ at the excision region to
zero at the event horizon (EH), and is zero for $r\geq r_{\rm EH}$,
so that the
exterior of the BH is causally disconnected from the effect of this extra
term. $U_0$ is set to zero or to the value of the atmosphere if the
corresponding field has one.
The coefficient ${(\Delta x)}^p$ ensures that the damping term converges
to zero and will not modify the convergence rate as long as one chooses $p$
to be greater than or equal to the order of convergence
of the code. In this work we adopt $p=4$.

These equations are implemented within the \had\ computational infrastructure which
provides distributed Berger-Oliger style adaptive mesh refinement
(AMR)~\cite{had_webpage,Liebling} with full subcycling
in time, together with a novel treatment of artificial boundaries~\cite{Lehner:2005vc}.
Because of the dynamics involved in this work, it is only necessary to use a fixed
refinement hierarchy, covering with finer grids the (central) region containing the disk and black
hole, and increasingly coarser grids in the outer regions in order to locate
the boundaries far away at a low computational cost.

%
%

\section{Overview of the physical setup:}
To explore the effects of the black hole merger in the dynamics of the accretion disk,
we concentrate, in particular, in the post-merger stage
--when the main burst of gravitational radiation has passed through the disk
and this has settled down to a quasistationary state--.\footnote{Studies of possible premerger
effects are presented in, e.g.,~\cite{Palenzuela:2009yr,chang}.}
To simulate a BH formed through the merger process and account
for the main effects of mass loss or recoil, we either consider a reduction in the
mass of the black hole by $5\%$ or apply a boost to the BH in a given direction.
In the latter case, it is easier to adopt the BH's rest frame and
apply the boost to the fluid variables (in the opposite direction) describing the disk,
which is represented by a stationary toroidal solution of the fluid equations in a Kerr background.
Thus, starting with a stationary
torus on a Kerr background, we perform a Lorentz boost with velocity $-\vec{v}_{\rm kick}$ on the
disk. We employ this  boost to transform the fluid's 4-velocity $u^\mu$ and magnetic field
4-vector $b^\mu$ when considering the recoil case.

The toroidal solutions are constructed following an
approach similar to that in \cite{torus.shibata}, adapted to the ingoing
Kerr-Schild coordinates adopted in our studies, and with a different
choice of specific angular momentum for the fluid for easier comparison with
previous work in the absence of magnetic fields . While in the current work we
do not simulate scenarios that include a magnetic field, we discuss the construction
of initial data that allows for doing so for future reference.
 In our case, we adopt the more standard
$l \equiv  -{u_{\phi}} \, (u_t)^{-1}={\rm const.}$ (see below) to allow for an
easier comparison with
previous work in the absence of magnetic fields. In particular, we verify that
identical solutions to those of \cite{AJS} are obtained if the magnetic
field is set to zero. In what follows we review the main steps in this construction.


The stress-energy tensor for ideal MHD can be written as
\begin{equation}
 T_{\mu \nu} = \left(\rho h + b^2 \right) u_{\mu} u_{\nu}
           + \left( P +  b^2/2 \right) g_{\mu \nu}
           - b_{\mu} b_{\nu},
\end{equation}
where $\rho$, $P$, and $u^\mu$ are the fluid's
density, pressure and 4-velocity, respectively, $b^\mu$ is the
magnetic field 4-vector, and $h$ is the specific enthalpy, defined as
\begin{equation}
 \rho h = \rho (1+\epsilon) + P,
\end{equation}
where $\epsilon$ is the specific internal energy density.

For the construction of initial data, we work with cylindrical coordinates
$(t,r,\phi,z)$ and
 make the assumption that the space-time is stationary and
axially symmetric. We adopt coordinates adapted to these symmetries, so
that only the $t$- and $\phi$-components of $u^{\mu}$
and
$b^{\mu}$ are nonzero.


The fluid equations are obtained from
\begin{equation}
  \nabla_{\mu} {T^{\mu}}_{\nu} = 0, \label{divT}
\end{equation}
together with the continuity equation ${\nabla_{\mu}(\rho u^{\mu})=0}$ (which
is trivially satisfied under our
assumptions). After some manipulation,
equation~(\ref{divT}) can be reduced to the integral equation
\begin{eqnarray}
&&\int\! u^t u_\phi \, d\left(\frac{u^\phi}{u^t}\right) - \ln u^t \nonumber \\
& &~~~~~~~ +
  \int\! \frac{1}{h \rho}\; dP
  + \int\! \frac{1}{2 \rho h D}\; d(b^2 D) = {\rm const.},
\end{eqnarray}
where $D=|g_{t t}g_{\phi \phi}-g_{t \phi}^2|$.  This equation
can be integrated after imposing further conditions that fix relationships
between the fluid variables as discussed below.

First, we fix a relationship between the velocity components. This can
be accomplished by requiring that the specific angular momentum $l$ satisfies
\begin{equation} \label{eq:defL}
 l \equiv  -\frac{u_{\phi}}{u_t} = {\rm const.}
\end{equation}

Second, we assume an isentropic fluid, imposing \(dh=\rho^{-1}dP\), which allows us
to integrate one of the terms out. An equation of state that satisfies this condition is
that of a polytrope
\begin{equation} \label{eq:poly}
P=\kappa \rho^{\Gamma} \, .
\end{equation}
In this case, the specific internal energy density can be calculated as
\begin{equation} \label{eq:epsilon}
\epsilon = \frac{\kappa}{\Gamma-1} \rho^{\Gamma-1} .
\end{equation}

We adopt this condition only to obtain the stationary
solutions for initial data. The fluid's entropy will change after the kick and
so we adopt, during the evolution, a \noindent{$\Gamma$-law} equation of state
\begin{equation}
P=(\Gamma-1)\rho \epsilon
\end{equation}
with $\Gamma=5/3$ considering the gas as being monoatomic.

Finally, we impose a convenient expression for $b^2$ in terms of
other variables to integrate the last term
\begin{equation} \label{eq:defB}
b^2\;D  = C  {( \,\rho h \;D)}^q  ,
\end{equation}
where  $C$ and $q>1$ are arbitrary constants.

After integrating eqn. (6), we use (\ref{eq:poly}) and (\ref{eq:epsilon})
to eliminate $\rho$ and $\epsilon$ and obtain an algebraic equation for $P$, of the
form
\begin{equation} \label{eq:implicitP}
F(P,g_{\mu \nu}, l, C, q) =  F_0,
\end{equation}
where $F_0$ is a constant of integration.
This equation can be solved analytically in the absence of magnetic field ($b^2=0$),
otherwise a straightforward numerical integration can be set up to obtain the solution.
The boundary of the torus is determined by setting $P=0$, obtaining an
expression of the form
\begin{equation}
f(g_{\mu \nu}, l) = F_0,
\end{equation}
which, through the dependence of $g_{\mu \nu}$ on the coordinates, is an
implicit surface equation. Notice that it is independent of both $C$ and $q$ so
that the location of the disk's boundary is independent of the magnetic
field. The solutions obtained may be toroidal as well as spheroidal, depending on
the values of $l$ and $F_0$.

Once $P$ is known, one can use once again equations (\ref{eq:poly}) and
(\ref{eq:epsilon}) to recover $\rho$ and $\epsilon$. The
velocity $u^\mu$ is obtained from equation (\ref{eq:defL}) together with the
normalization condition $u^\mu u_\mu=-1$. Finally, the magnetic field $b^\mu$
is determined by equation (\ref{eq:defB}) together with the relation
$b^\mu u_\mu = 0$ (see \cite{had1}).
The magnetic filed is always zero at the surface of the disk (from equation
\ref{eq:defB}), and one can control how rapidly it decays to zero with the
parameter $q$, and its maximum magnitude with $C$.

\begin{figure}[ht!]
\begin{center}
\includegraphics[angle=0, width=0.49\columnwidth,clip]{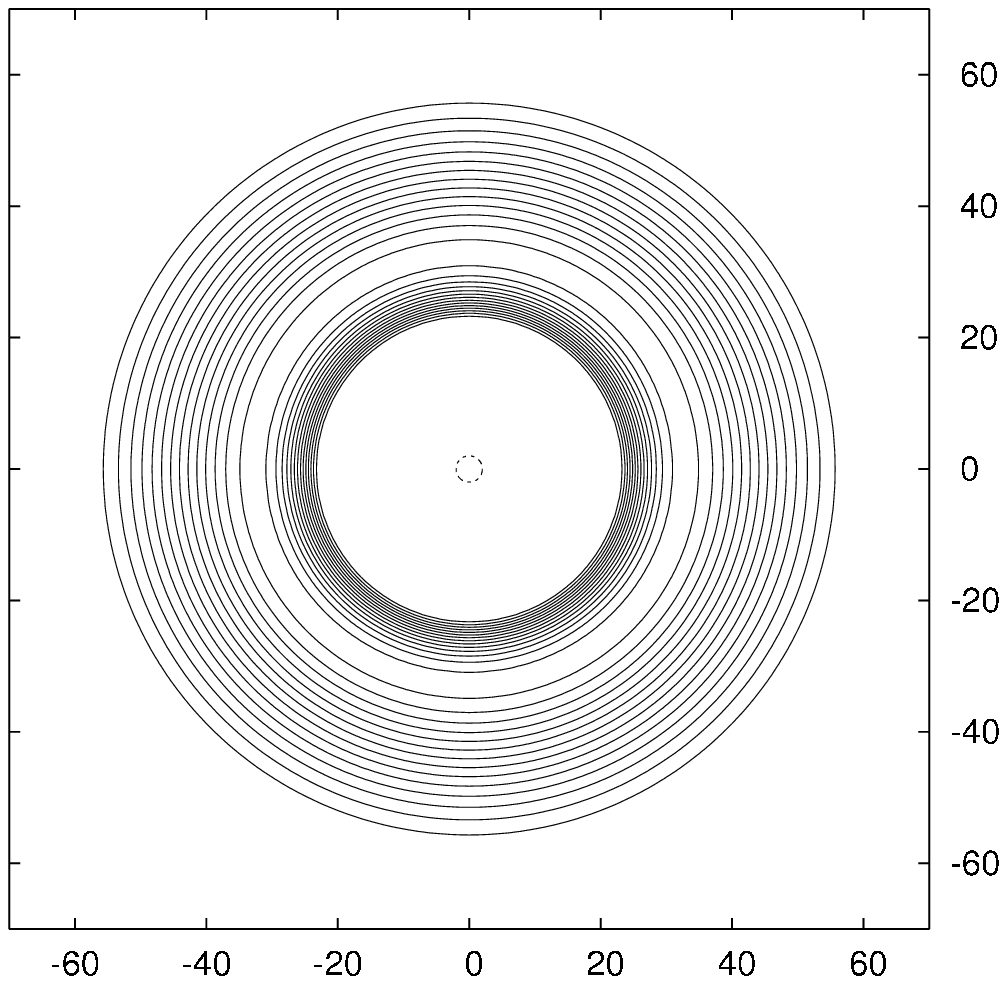}
\includegraphics[angle=0, width=0.49\columnwidth,clip]{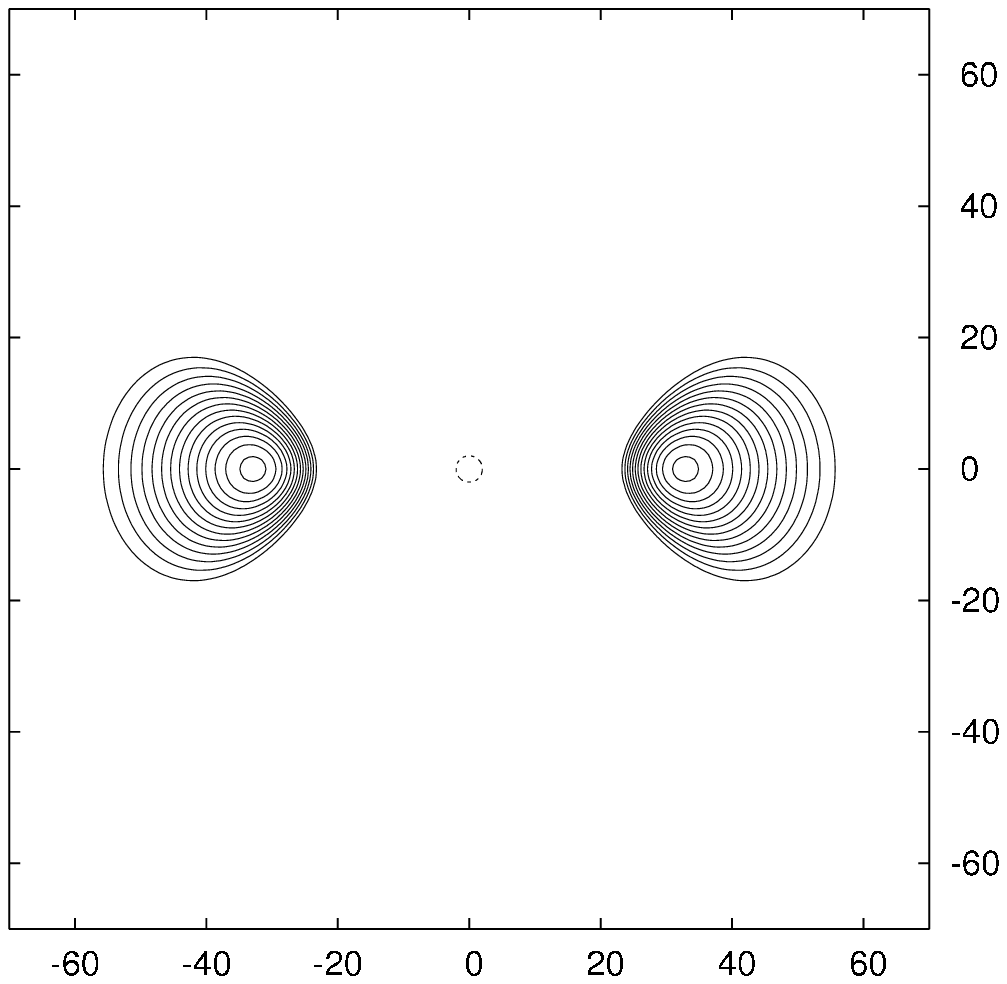}
\caption{Representative example of the toroidal initial configurations, showing the
  density at the equatorial plane (left panel),
  and at a meridional plane (right panel). The dashed line indicates the
  location of the event horizon. \label{f:DVrho0}}
 \end{center}
\end{figure}

As mentioned, when considering the recoiling case, 
the initial data for a disk is given a Lorentz boost
with respect to the stationary system of the background black hole.
A representative example of the
toroidal configurations is
shown in Figure~\ref{f:DVrho0}. Such configurations are then
evolved on a computational domain given by
$[-150 \,M,150 \,M]$ in the $x$- and $y$-direction, and $[-100 \,M,100 \,M]$
in the $z$-direction (since the disk lies on the $x-y$ plane), with an  FMR
configuration having 3 levels of refinement.
The code used in our studies has been previously tested and employed in a variety of
stringent scenarios, e.g.,\cite{had2,binaryNS,Anderson:2008zp}. For our specific
application we have verified that in the absence of a kick or mass reduction
the disk remains stationary during the evolution as expected. Additionally,
we have verified convergence by comparing results obtained with three different
resolutions in the case of a kick velocity of 3000~km/s perpendicular to the
axis of rotation. The convergence rate measured at different locations
varies between first and third order depending on the presence of shocks.

Certainly the parameter space is too vast to allow an exhaustive computational
study. Therefore,
we mainly concentrate here on varying the most relevant parameters, i.e., the kick
magnitude and direction
and study a few other cases varying the spin parameter to verify
our results are qualitatively the same. Notice that variations with respect to the
spin parameter $a$ should not lead to significant qualitative differences unless
accretion develops, as
the disk's inner edge is located sufficiently far away for its influence to
be of higher order. This intuitive observation is confirmed by our simulations.

The toroidal solutions employed in this work correspond to specific angular momentum $l/M=6$,
spin parameter $a/M=0.5$ (except when analyzing the solution's dependence on the spin
where we also consider $a/M=0.9$). Also, we fix the magnetic field parameter
$C=0$ (so that $b=0$), and choose $F_0$
so that the inner edge of the disk is located at $r_{\rm in}=20M$. With this choice of
parameters, the outer edge is located at $r_{\rm out}=60M$ and the maximum
pressure in the disk lies at $r_{\rm m}=33M$. The orbital velocity of the fluid is then
$0.28, 0.17$ and $0.10$c at
$r_{\rm in},r_{\rm m}$ and  $r_{\rm out}$ respectively. Thus, the orbital period
at $r_{\rm m}$ is $P_{\rm m}=1220M$. The sound speed
has a maximum value 
$\lesssim 0.05$c close to $r_{\rm m}$, and drops abruptly to zero at the boundary of
the torus. All fluid elements in the torus
have an orbital speed much greater than the highest kick velocity adopted in this
work, i.e., 0.01c=3000~km/s
and so will remain bound to the black hole in all cases considered\footnote{For comparison
purposes we have also employed the unrealistic value of 9000~km/s}.
In fact, the binding energy per unit mass at the surface of the torus is $0.0121 c^2$,
which implies a escape velocity of $0.155c$.\\
Notice that the location of the disk's inner radius
can vary significantly depending on diverse physical parameters (e.g. kinematic
viscosity of the gas, accretion rate, binary mass ratio, etc.)~\cite{Milosavljevic:2004cg}.
We adopt a small value but within the allowed ones to reduce the computational cost
of the long simulations required and concentrate on extracting physically robust
conclusions, which can be intuitively extended to general cases.\\

Throughout the rest of this paper, unless otherwise specified, all kick
orientations mentioned refer to the kick (or Lorentz boost) applied to
the disk, which would correspond to the black hole being kicked in the
opposite direction.

%
%
\section{Results:}
\subsection{Diagnostic quantities:}
We monitor the fluid's behavior by examining the dependence of
the primitive values as different physical parameters are varied.
Ultimately, our goal is to understand possible electromagnetic signals
emitted by the system as the disk' dynamics is affected. At present, our
simulations do not incorporate radiation transport; thus, a direct computation
of these signals is not possible. Therefore, we concentrate on related quantities,
which when combined with a suitable model, can be tied to possible emissions.
In particular, we compute (an approximation to the) temperature ($T$),
the total internal energy ($U$) and bremsstrahlung luminosity ($L_B$) as
\begin{eqnarray}
T &\propto& P/\rho \, ,\\
U &\propto&\int \rho \epsilon dV \, ,\\
L_B &\propto& \int \rho^2 T^{1/2} dV \, .
\end{eqnarray}
Notice that unless the disk is optically thin, the bremsstrahlung luminosity need not capture the
luminosity resulting from shocks and shock heating. While the bremsstrahlung luminosity is
a good measure of the energy exchanged between atoms and the radiation field, it does not
take into account how this energy can be radiated out of the disk. In the absence of a more
refined model, the qualitative features of the
true radiative behavior can be estimated simply by a black body assumption. We thus
monitor the internal energy for this purpose and also the bremsstrahlung luminosity to
obtain a measure of the mentioned energy exchange (as well as to make contact with results presented
in \cite{massloss}).

\subsection{Axisymmetric cases: Black hole mass loss and kick along disk's angular momentum}
As a first step we consider the effect of BH mass loss and that of a kick
along the disk's orbital angular momentum. The former entails
solely decreasing the mass of the black hole, while for the latter the mass is unchanged
but a kick is introduced along the $z$ axis. In both cases, the underlying axisymmetry of the problem is
not broken, which as we shall see later, is a key issue.\\

\begin{figure}[ht!]
\begin{center}
\includegraphics[angle=0, width=0.49\columnwidth,clip]{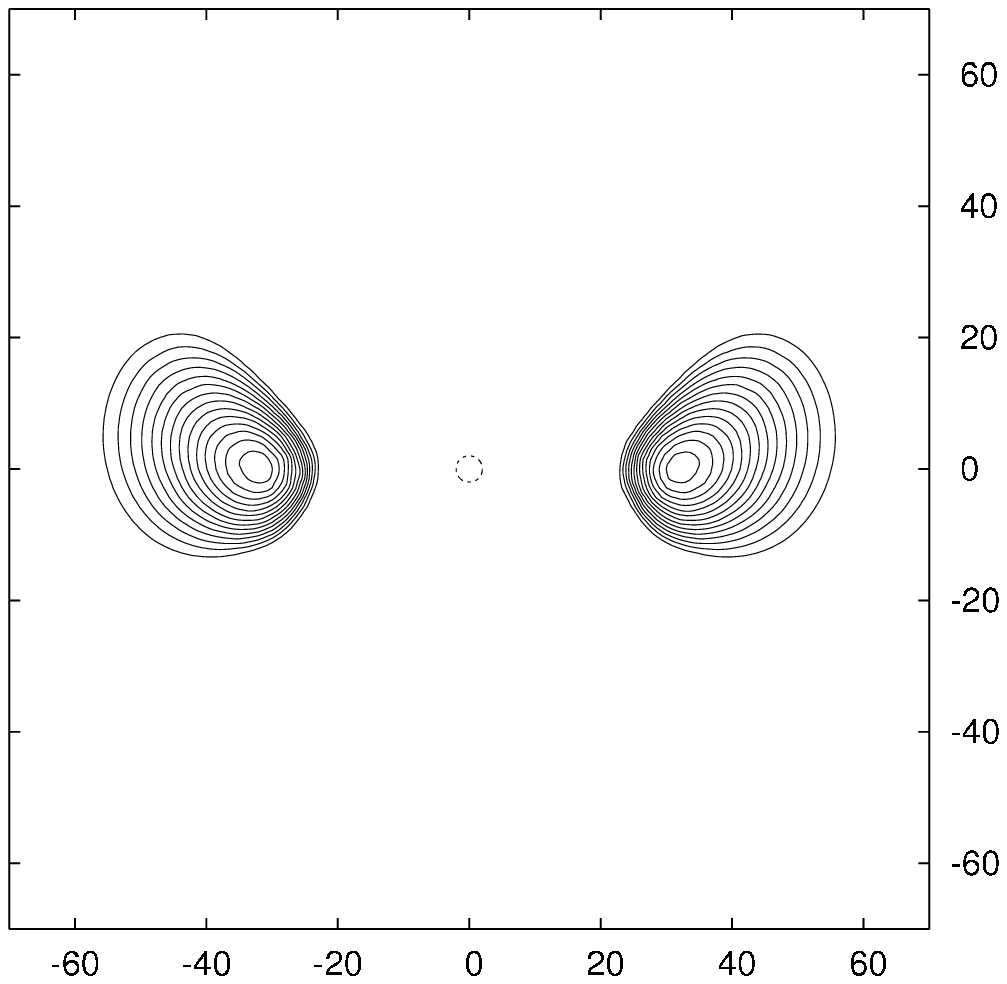}
\includegraphics[angle=0, width=0.49\columnwidth,clip]{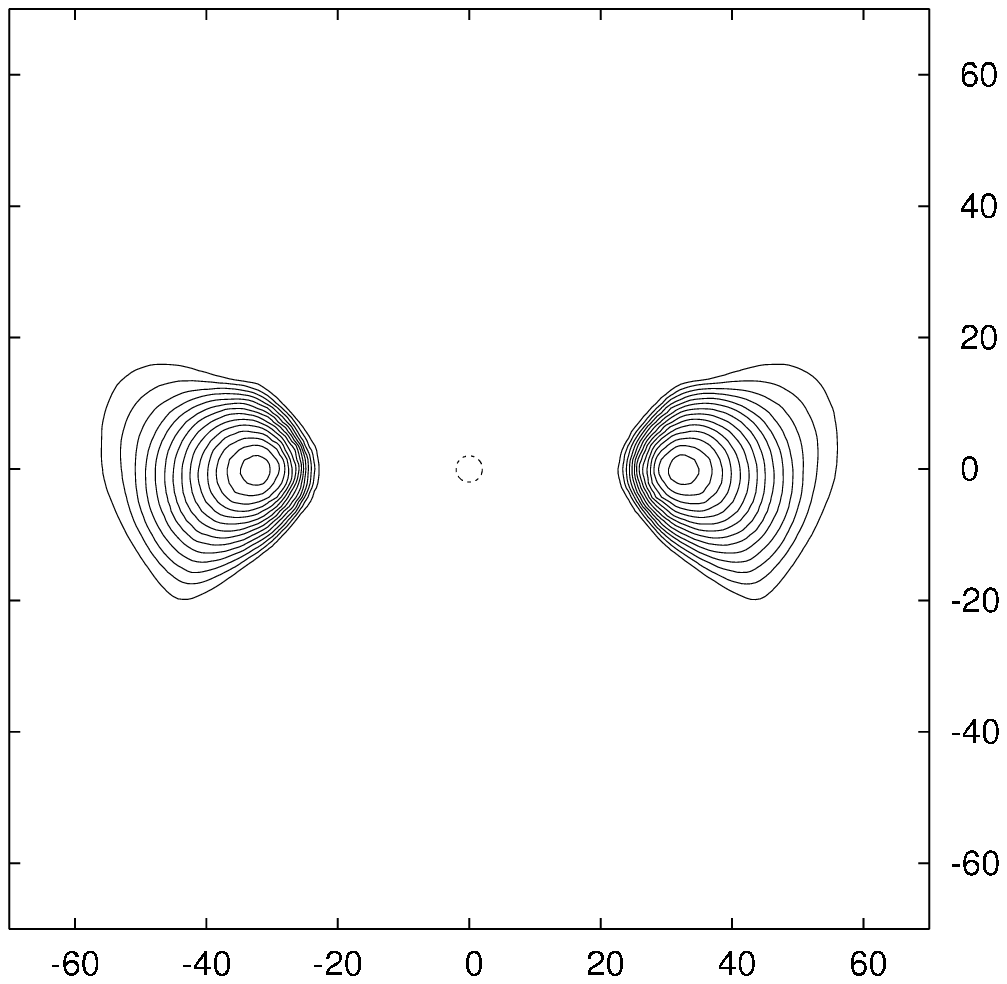}
\includegraphics[angle=0, width=0.49\columnwidth,clip]{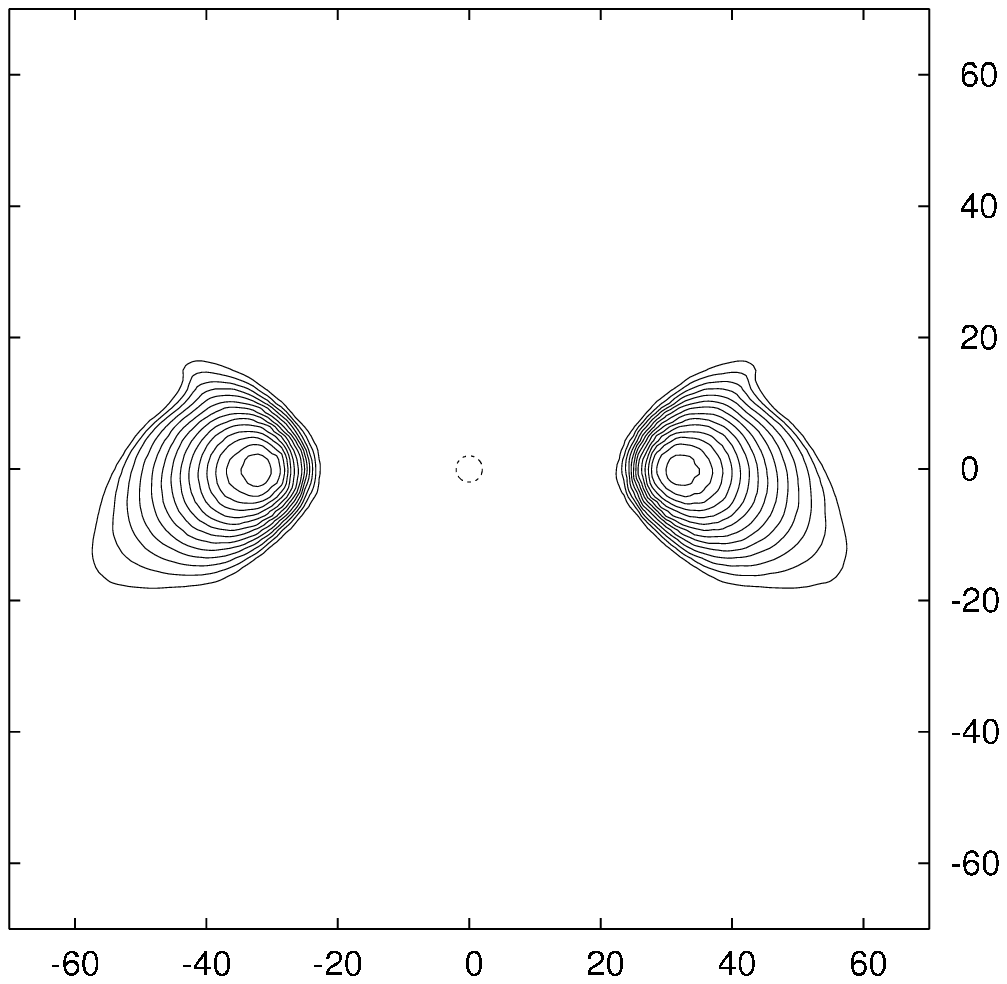}
\includegraphics[angle=0, width=0.49\columnwidth,clip]{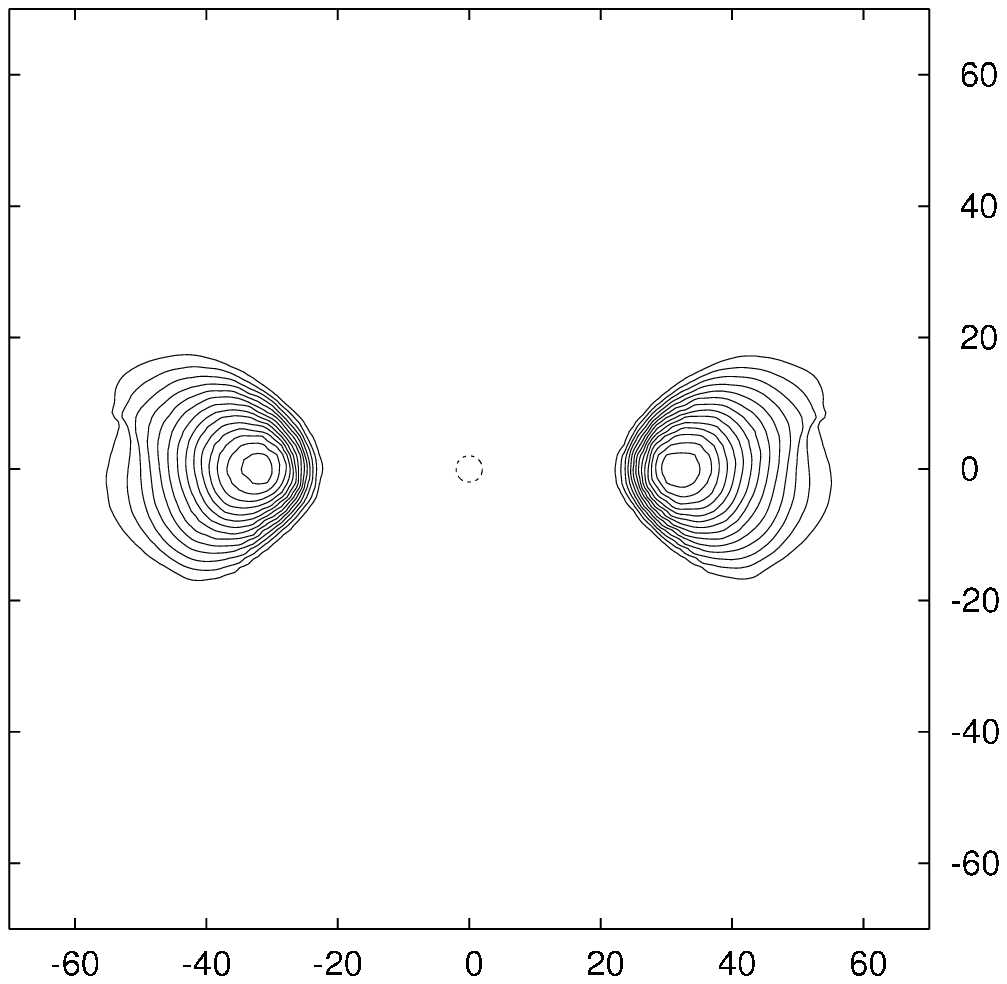}
\caption{Density at plane $y=0$ in the case of a disk kicked with a velocity
  3000~km/s in the positive z-direction. The panels show
  snaps from $t/M=500$ (top left) to $2000$ (bottom right) at $\Delta t/M=500$
  intervals.\label{f:rho3000a00} }
 \end{center}
\end{figure}

\begin{figure}[ht!]
\begin{center}
\includegraphics[angle=0, width=0.9\columnwidth,clip]{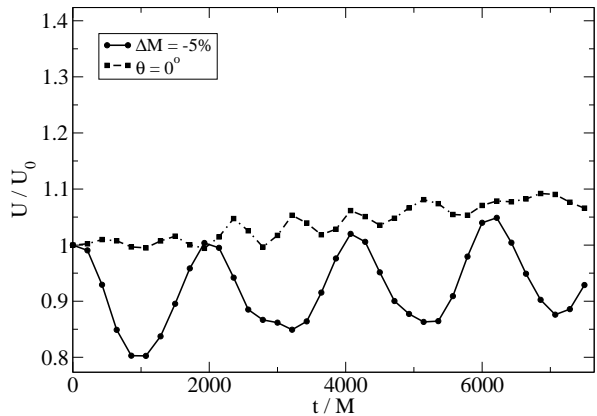}
\caption{Total internal (normalized) energy of the disk. The
  continuous line corresponds to a BH mass loss of
  5\% and no kick, while the dashed line corresponds to a
  kick with velocity $v_{\rm k}=3000$~km/s along the axis of rotation (and
  no BH mass loss). The vertical scale and range was chosen to coincide with
  those in the other energy plots in this work for easy comparison.\label{f:ie_mass_axis}}
 \end{center}
\end{figure}

\begin{figure}[ht!]
\begin{center}
\includegraphics[angle=0, width=0.9\columnwidth,clip]{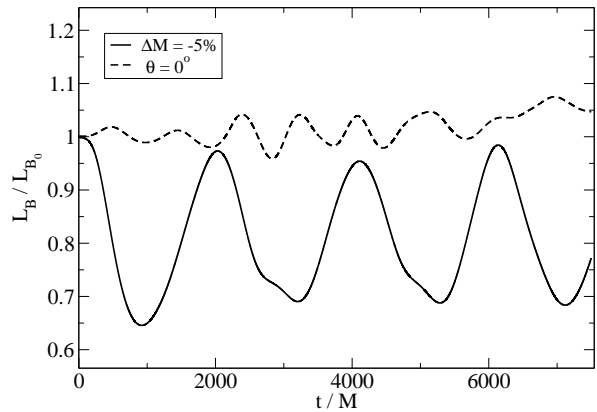}
\caption{Bremsstrahlung luminosity (normalized) of the disk. The
  continuous line corresponds to a BH mass loss of
  5\% and no kick, while the dashed line corresponds to a
  kick with velocity $v_{\rm k}=3000$~km/s along the axis of rotation (and
  no BH mass loss). \label{f:brem_mass_axis}}
 \end{center}
\end{figure}

For the mass loss case, we employ a toroidal solution corresponding to a
black hole of mass $M_0$ for the initial data, and set $M=0.95M_0$. The dynamics of the disk
with either a reduced mass or a kick along the $z$ axis behave in a rather smooth manner. For the
mass reduced case, radial oscillations are induced as the different fluid elements follow their
corresponding epicycles. For the case with a recoil velocity, further oscillations are
generated
by induced motions in the $z$ axis as illustrated in figure~\ref{f:rho3000a00}. Indeed,
the recoil motion of the black hole introduces a time-dependent vertical component
of the black hole's gravitational pull on the disk.
Using Newtonian mechanics for simplicity and ignoring pressure forces, one can show
that a particle on a circular orbit with velocity $v_{\rm orb}$, after a vertical kick 
of magnitude $v_{\rm kick}$ only reaches a height $z=\sqrt{2}R (v_{\rm kick}/v_{\rm orb})$
above the original plane before turning around. Since $v_{\rm kick}$ is the same for all
disk radii, the vertical displacement is minimal at $r_{\rm in}$ and maximal at $r_{\rm out}$. 
This results in a flexing axisymmetric mode, with the outer edge flopping about the most.
This is supported by Fig.~\ref{f:rho3000a00} if one defines the ``midplane" of
the disk by joining points at which the contours are vertical. This argument
ignores pressure, but pressure gradients are unlikely to be very important away from shocks,
and the behavior is qualitatively the same. Because all particles on a given radial annulus
are kicked simultaneously, they remain in phase with each other and the flexing mode is
 naturally excited. Note that maximum compression occurs twice per orbital period so this flexing mode 
is visible in both internal energy and bremsstrahlung at a frequency of about twice the orbital
(See Figs.~\ref{f:ie_mass_axis} and~\ref{f:brem_mass_axis}).

Most importantly, in either case no significant shocks are developed during the time
of these simulations ($\simeq 6 P_{\rm m}$ ). The observed smooth behavior translates
into a rather monotonic behavior in our diagnostic variables.
Figures~\ref{f:ie_mass_axis} and~\ref{f:brem_mass_axis} illustrate the
internal energy and bremsstrahlung luminosity, respectively.
The behavior observed in the latter case is qualitatively similar to results shown
in~\cite{massloss}, i.e., an initial drop followed by a recovery in
luminosity. Our simulations, which extend farther, indicate that this behavior continues
quasiperiodically.  Notice however that the disk geometry considered in~\cite{massloss} is
different from ours and the bremsstrahlung computed includes the inner portion
of the disk while we do so for the complete disk. 
Last, the small drift observed in figure~\ref{f:ie_mass_axis} is consistent with a
linear accumulation of numerical error. A similar linear drift is observed in
simulations of an unperturbed disk. This growth however is small --within $5\%$-- over
the length of the simulations considered ($\simeq 6 P_{\rm m}$ ) and significantly
smaller than the effects induced by the perturbations due to the recoiling black hole.

\subsection{Asymmetric cases: Kick with component orthogonal to disk's angular momentum}

\begin{figure}[ht!]
\begin{center}
\includegraphics[angle=0, width=0.49\columnwidth,clip]{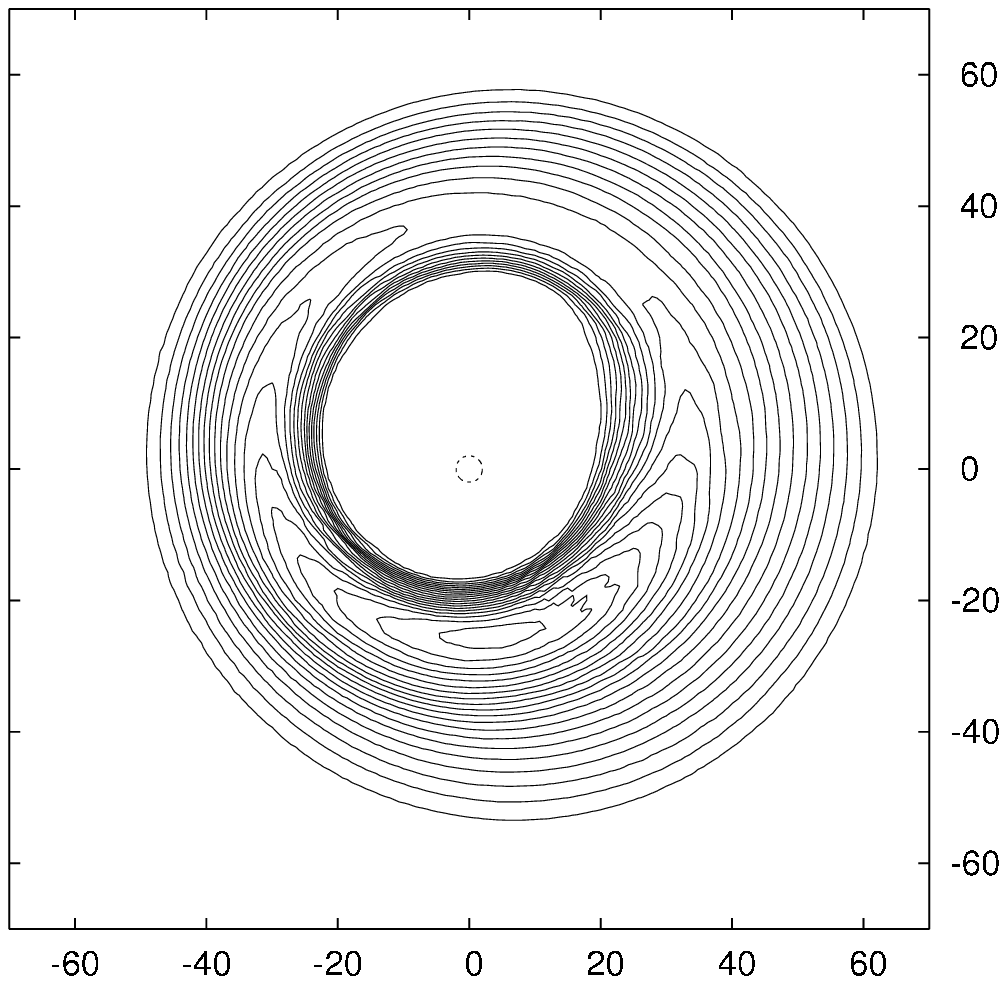}
\includegraphics[angle=0, width=0.49\columnwidth,clip]{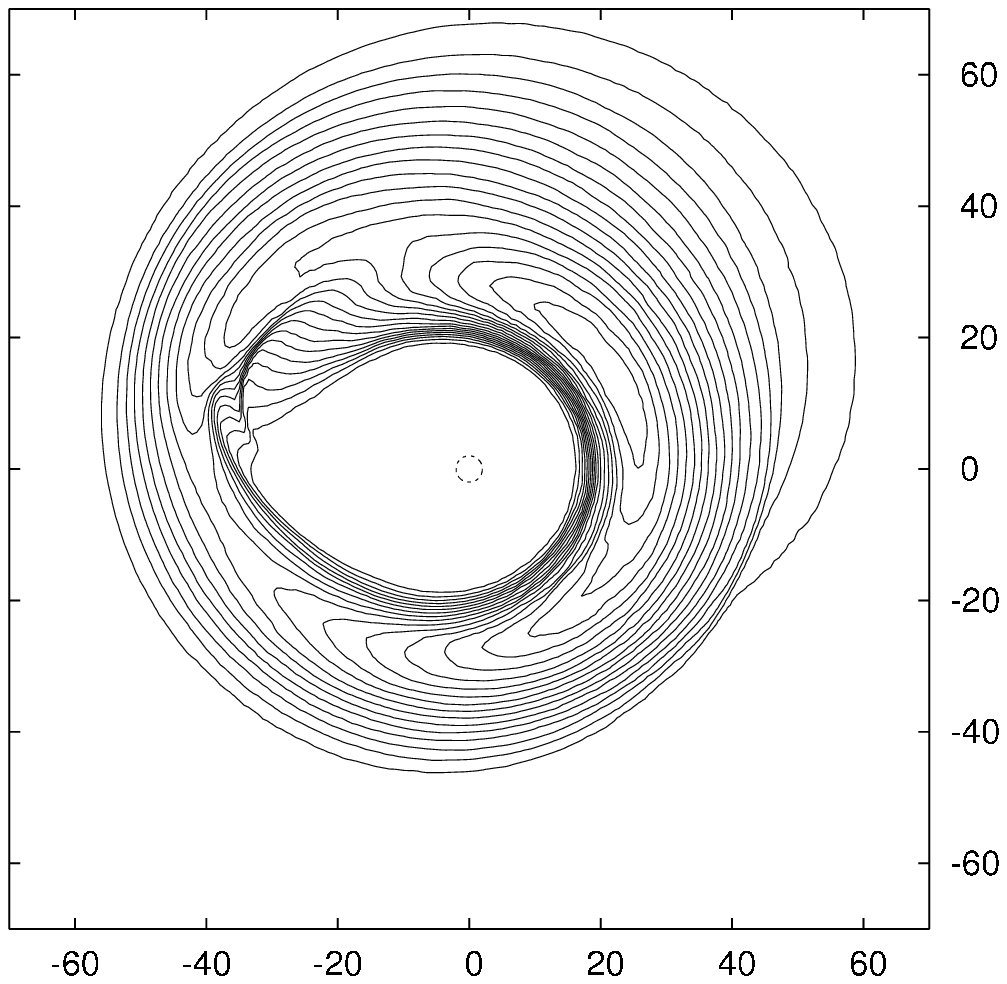}
\includegraphics[angle=0, width=0.49\columnwidth,clip]{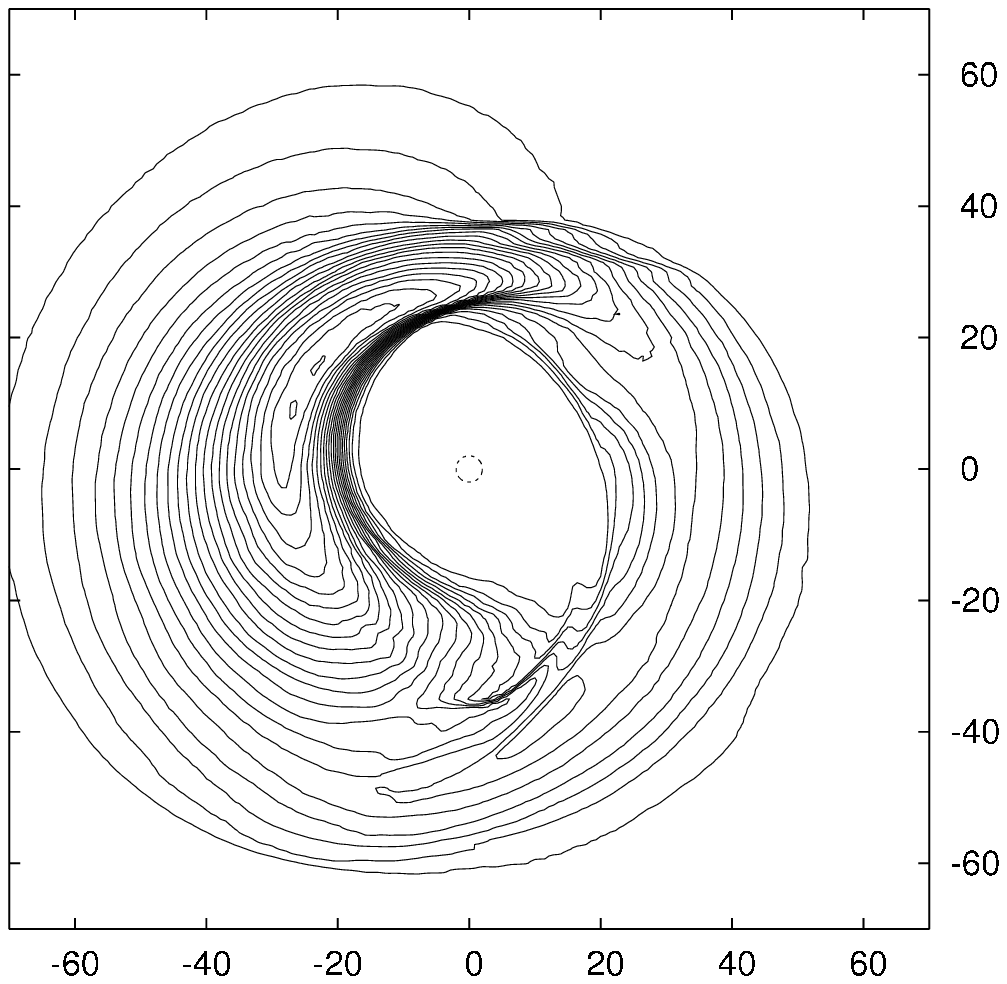}
\includegraphics[angle=0, width=0.49\columnwidth,clip]{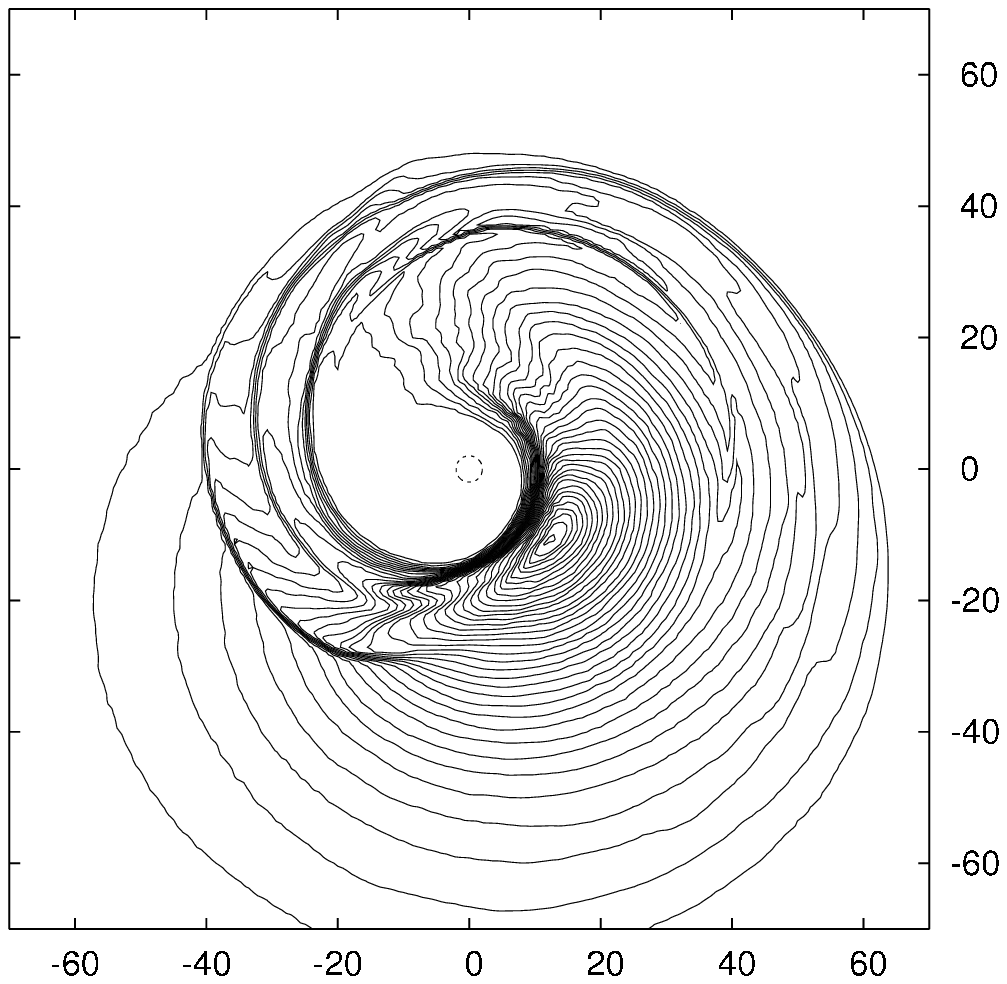}
\includegraphics[angle=0, width=0.49\columnwidth,clip]{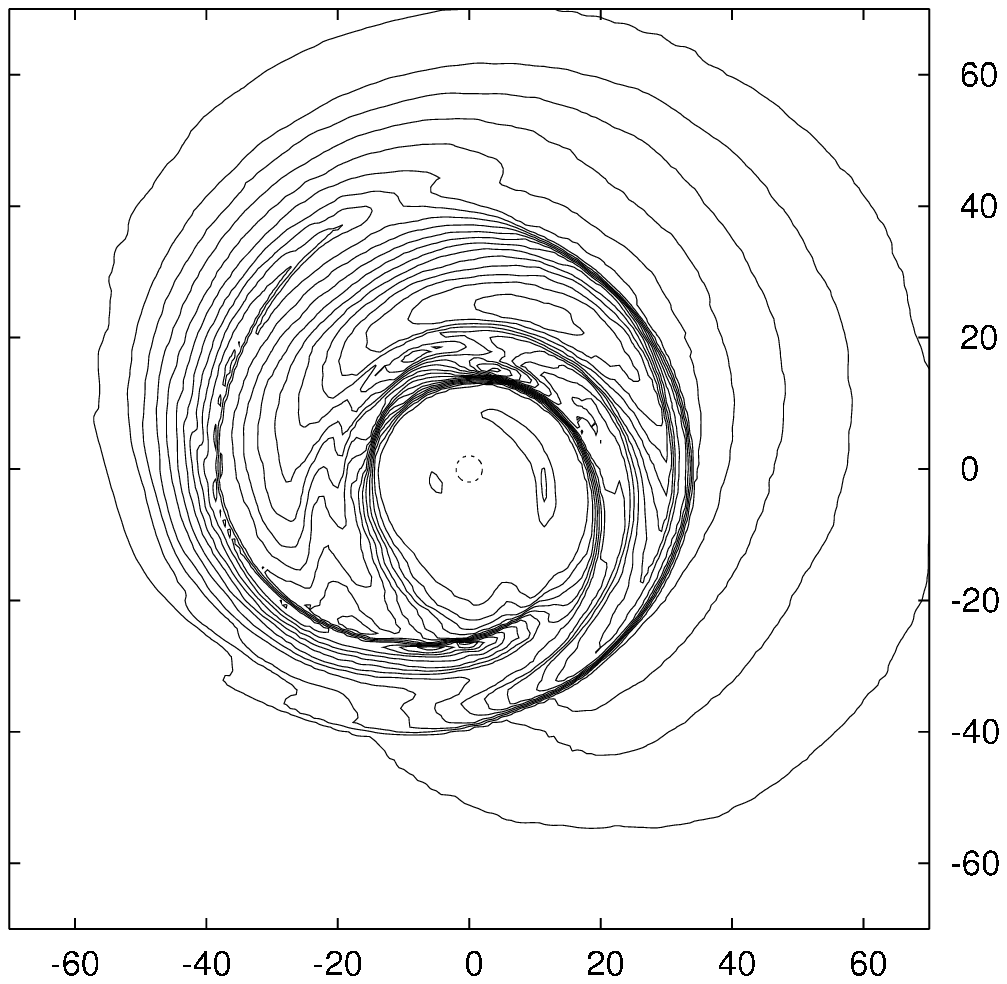}
\includegraphics[angle=0, width=0.49\columnwidth,clip]{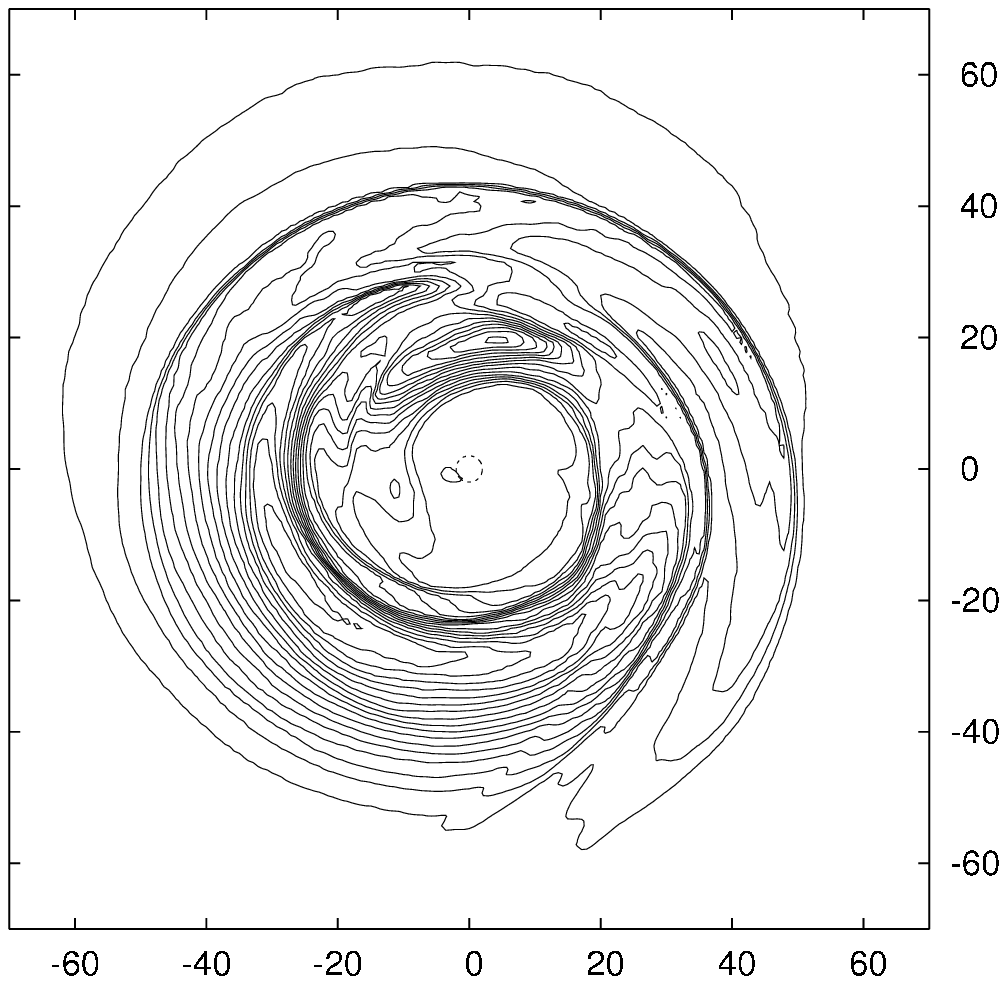}
\includegraphics[angle=0, width=0.49\columnwidth,clip]{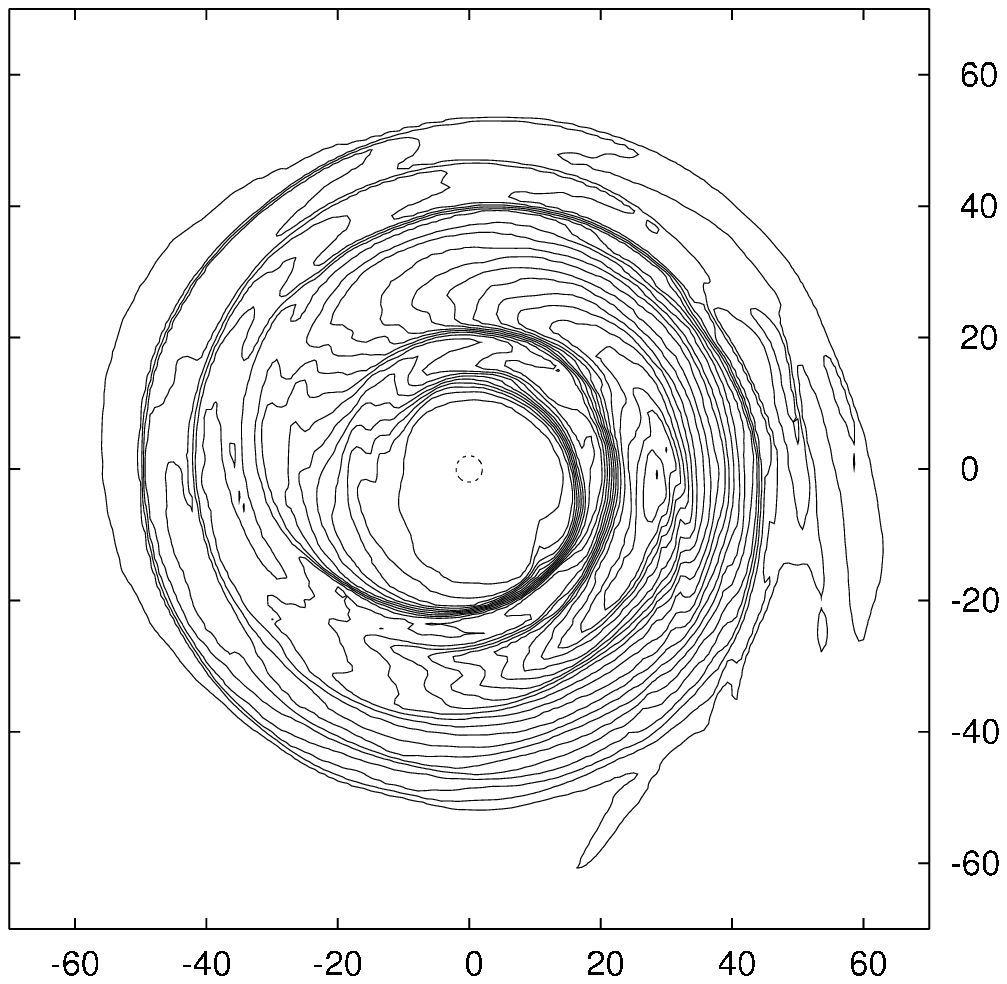}
\includegraphics[angle=0, width=0.49\columnwidth,clip]{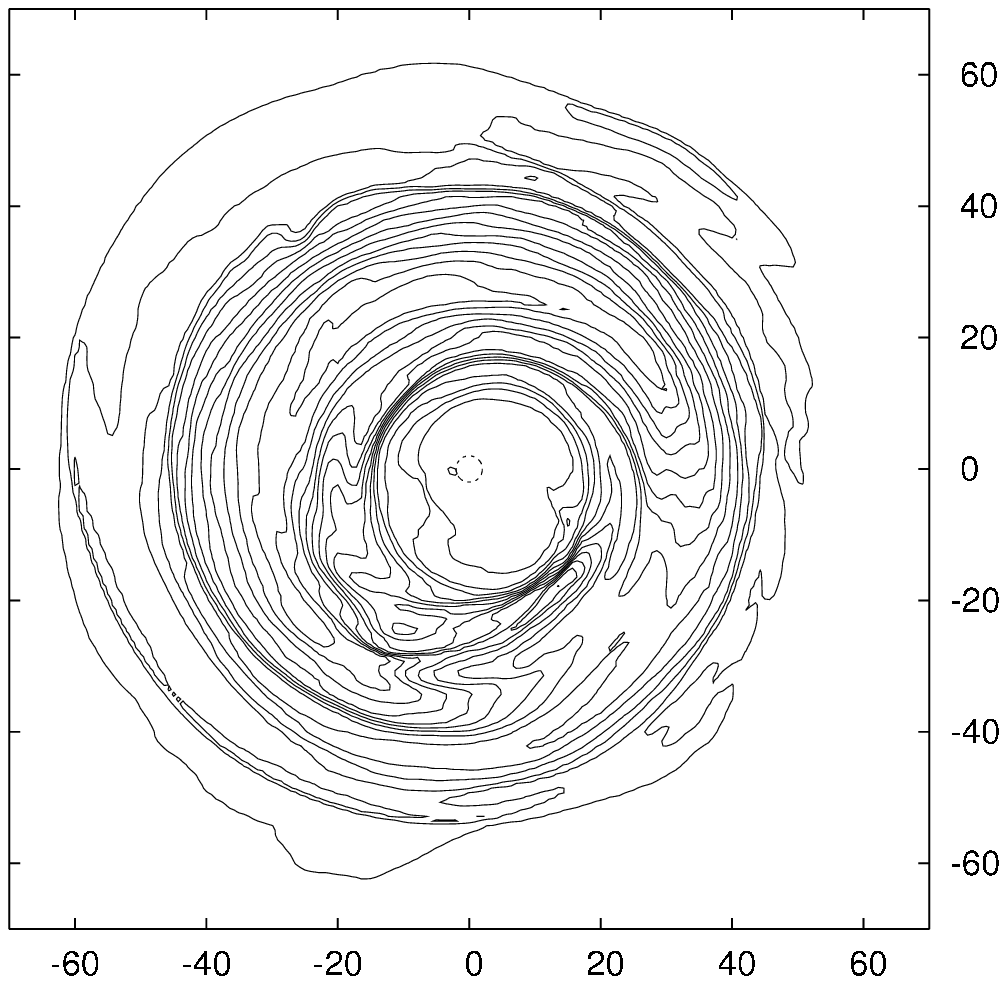}
\caption{Density $\rho$ at plane $z=0$ in the case of a disk kicked with a velocity
  3000~km/s in the positive x-direction, i.e., to the right of this page (which
  corresponds to the black hole being kicked to the left). The panels show
  snaps from $t/M=500$ (top left) to $4000$ (bottom right) at $\Delta t/M=500$ intervals.\label{f:DVrhoF3}}
 \end{center}
\end{figure}

\begin{figure}[ht!]
\begin{center}
\includegraphics[angle=0, width=0.49\columnwidth,clip]{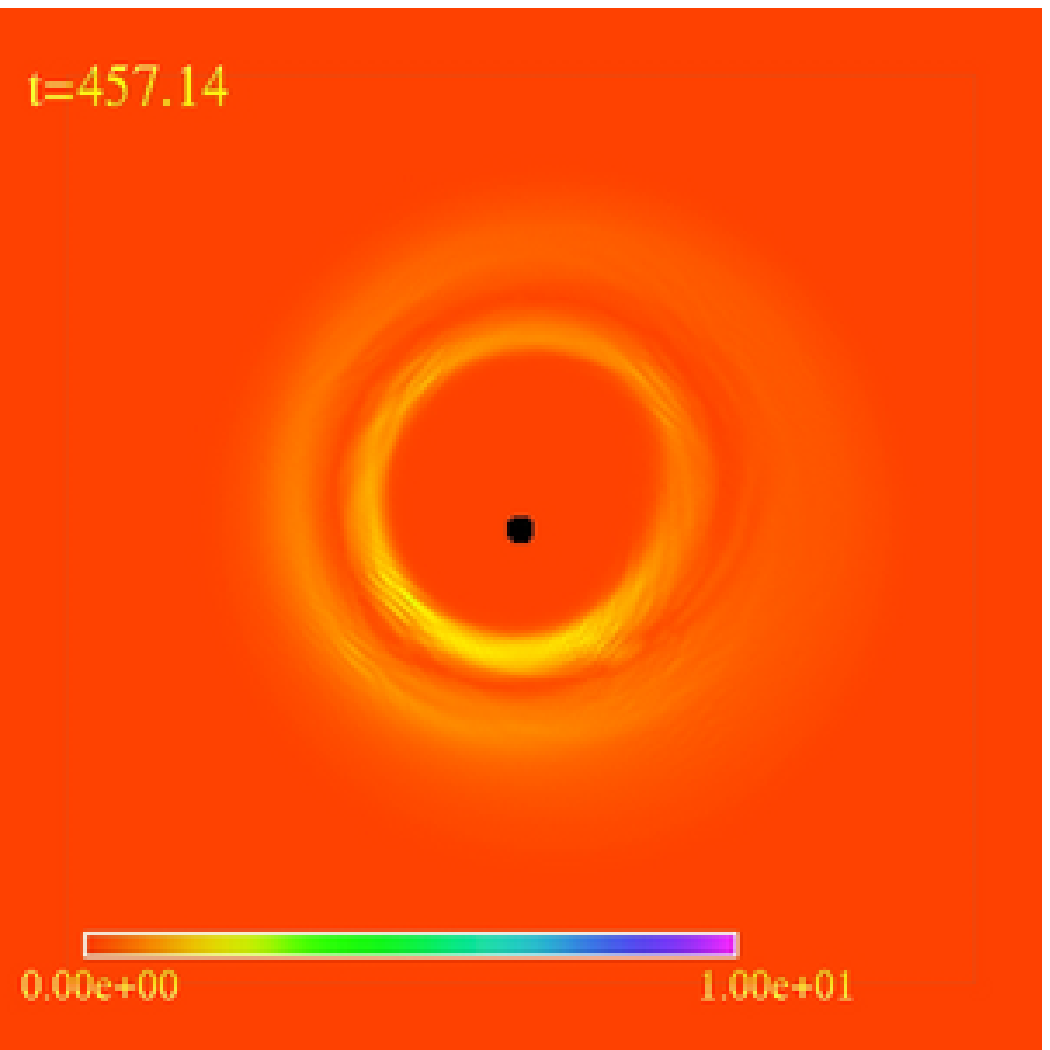}
\includegraphics[angle=0, width=0.49\columnwidth,clip]{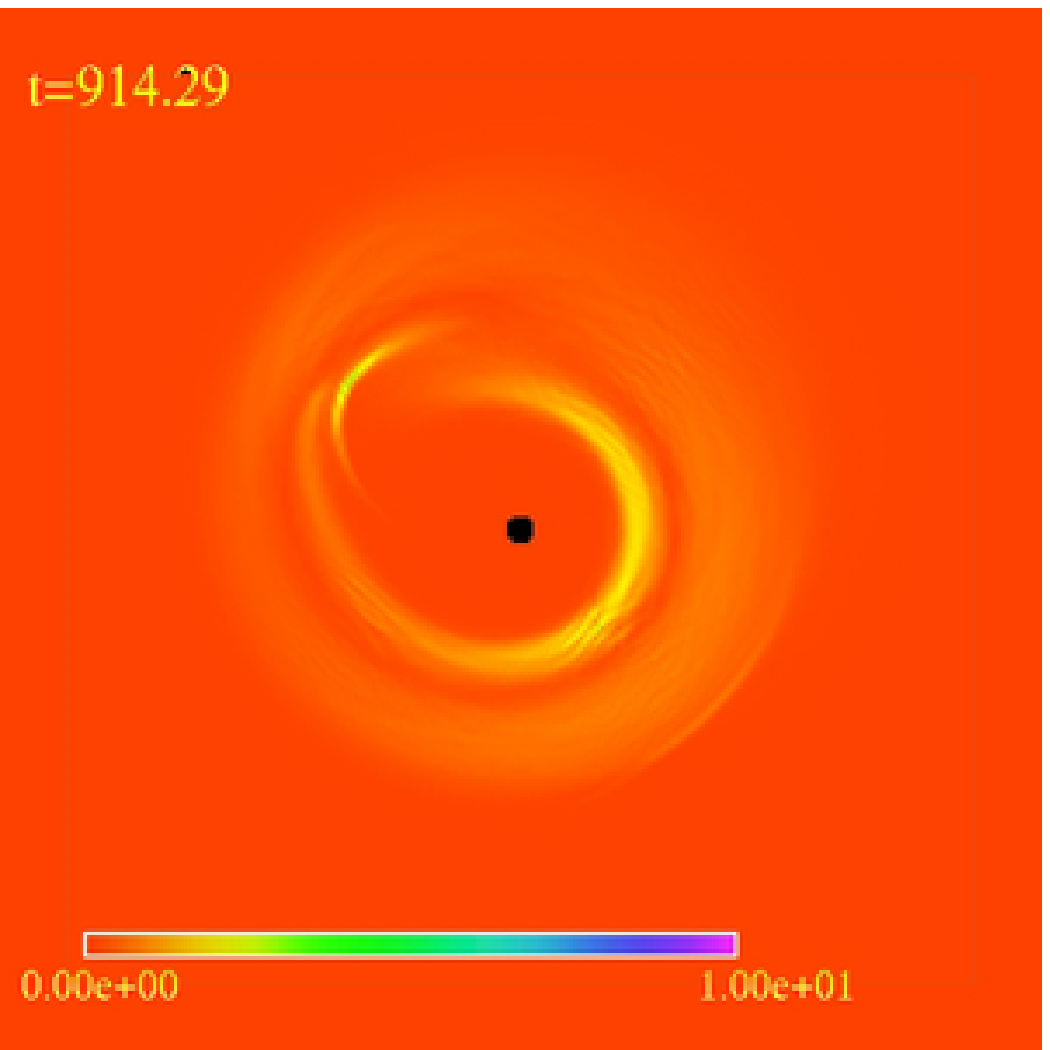}
\includegraphics[angle=0, width=0.49\columnwidth,clip]{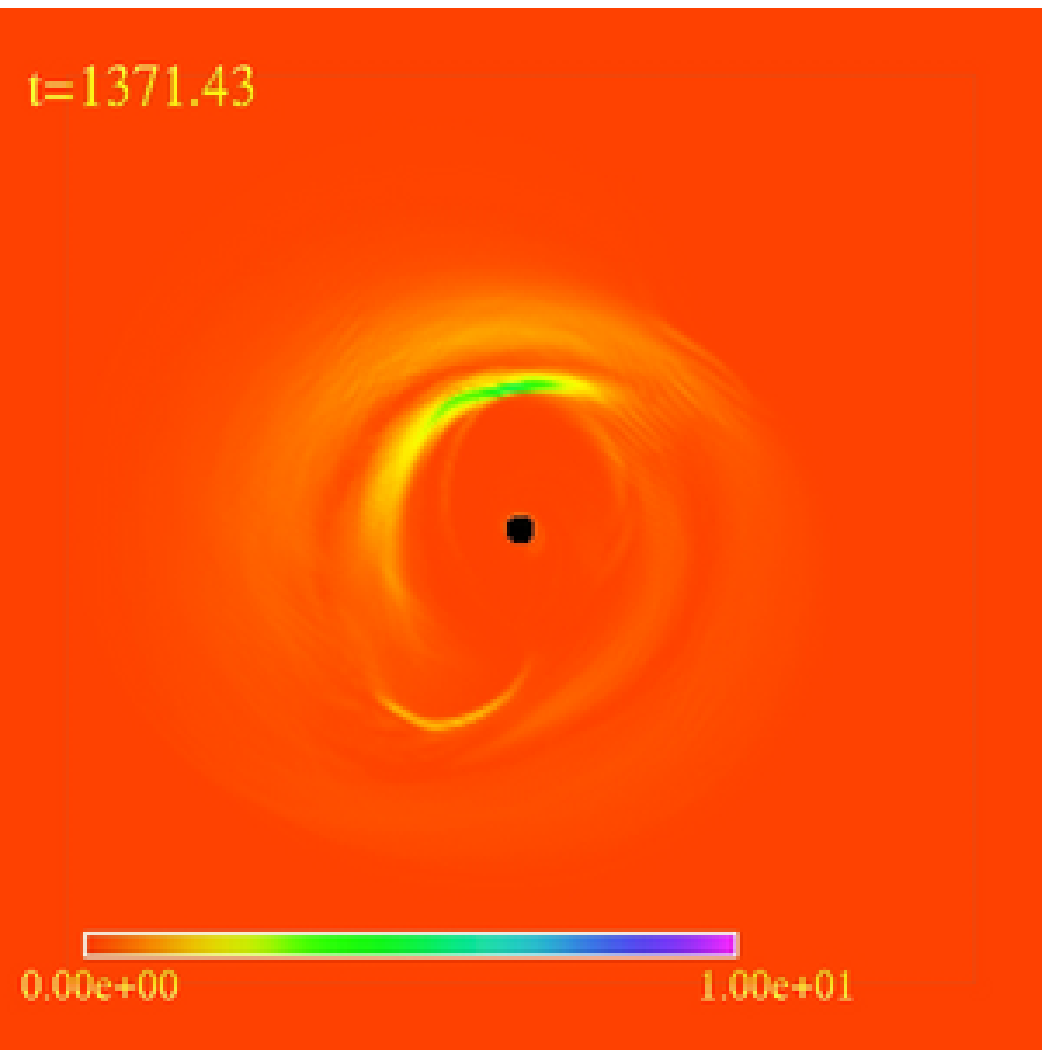}
\includegraphics[angle=0, width=0.49\columnwidth,clip]{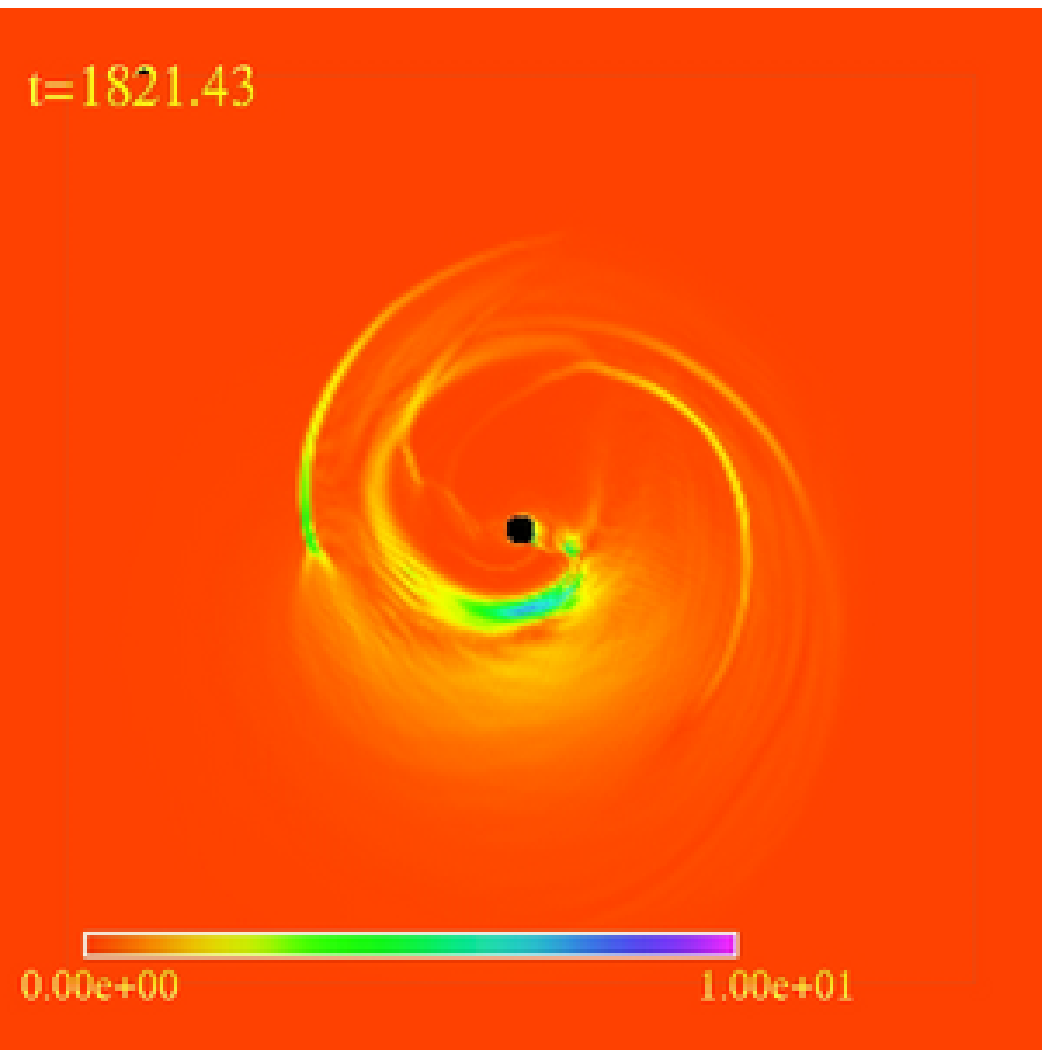}
\includegraphics[angle=0, width=0.49\columnwidth,clip]{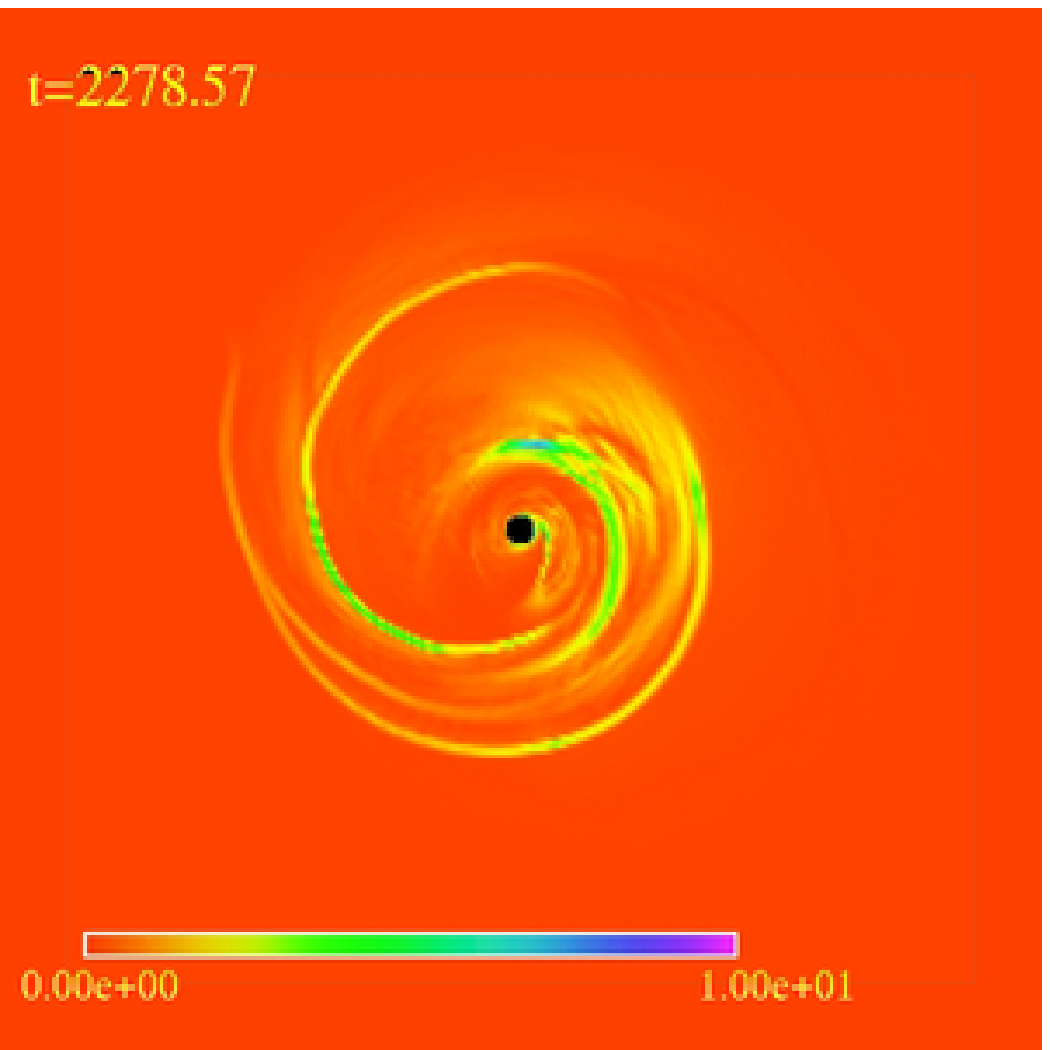}
\includegraphics[angle=0, width=0.49\columnwidth,clip]{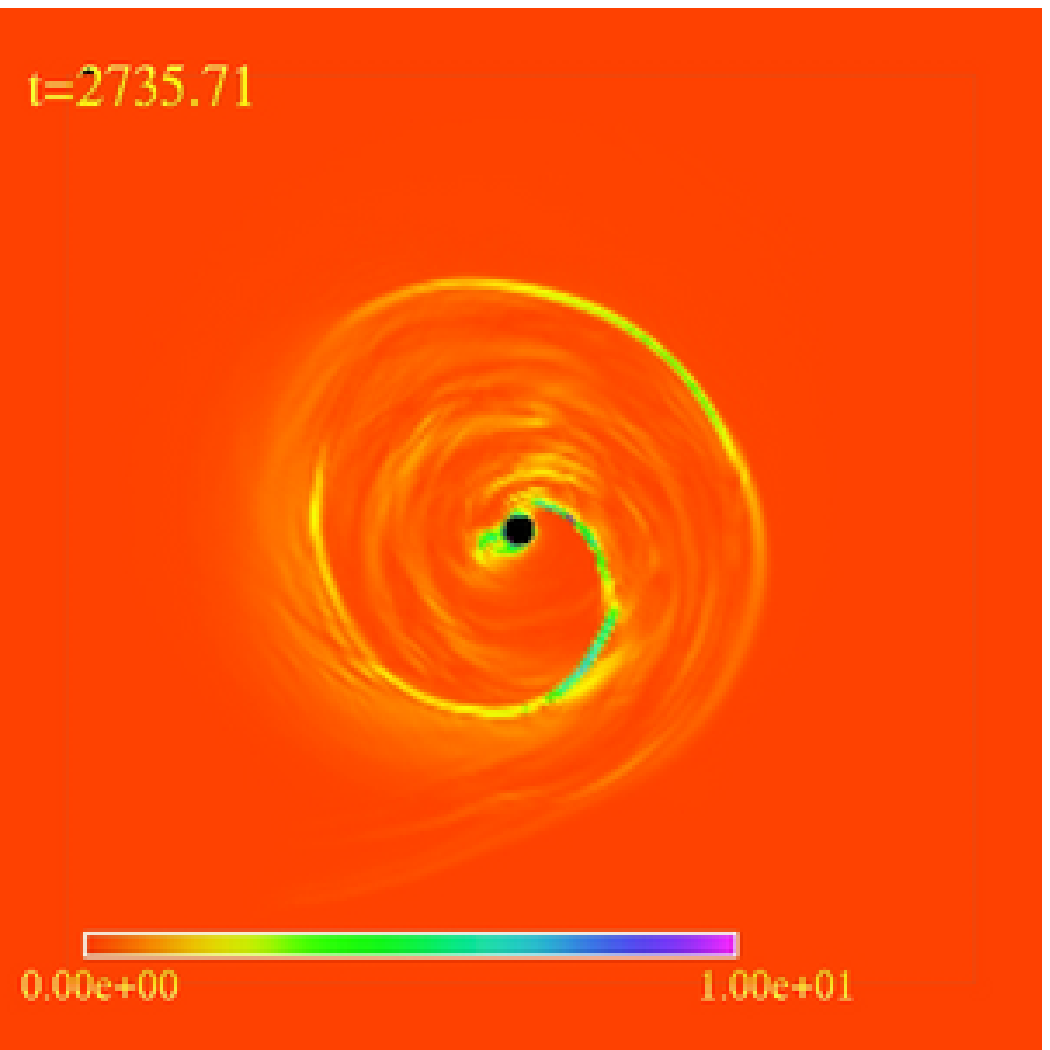}
\caption{To illustrate the formation of shocks we show here
  $|\vec{\nabla}P|$ at $z=0$. Kick velocity of 3000~km/s in the positive
  x-direction. \label{f:DVgradpF3}}
 \end{center}
\end{figure}

\begin{figure}[ht!]
\begin{center}
\includegraphics[angle=0, width=0.49\columnwidth,clip]{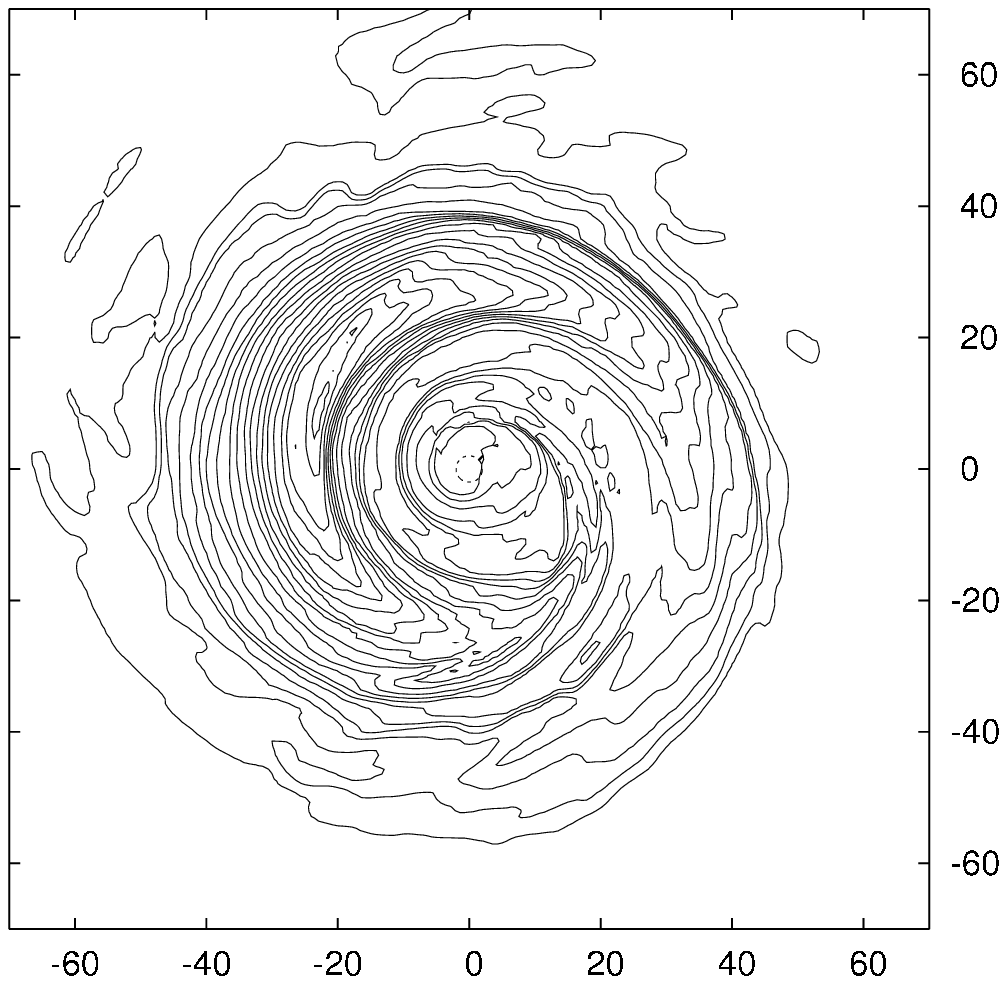}
\includegraphics[angle=0, width=0.49\columnwidth,clip]{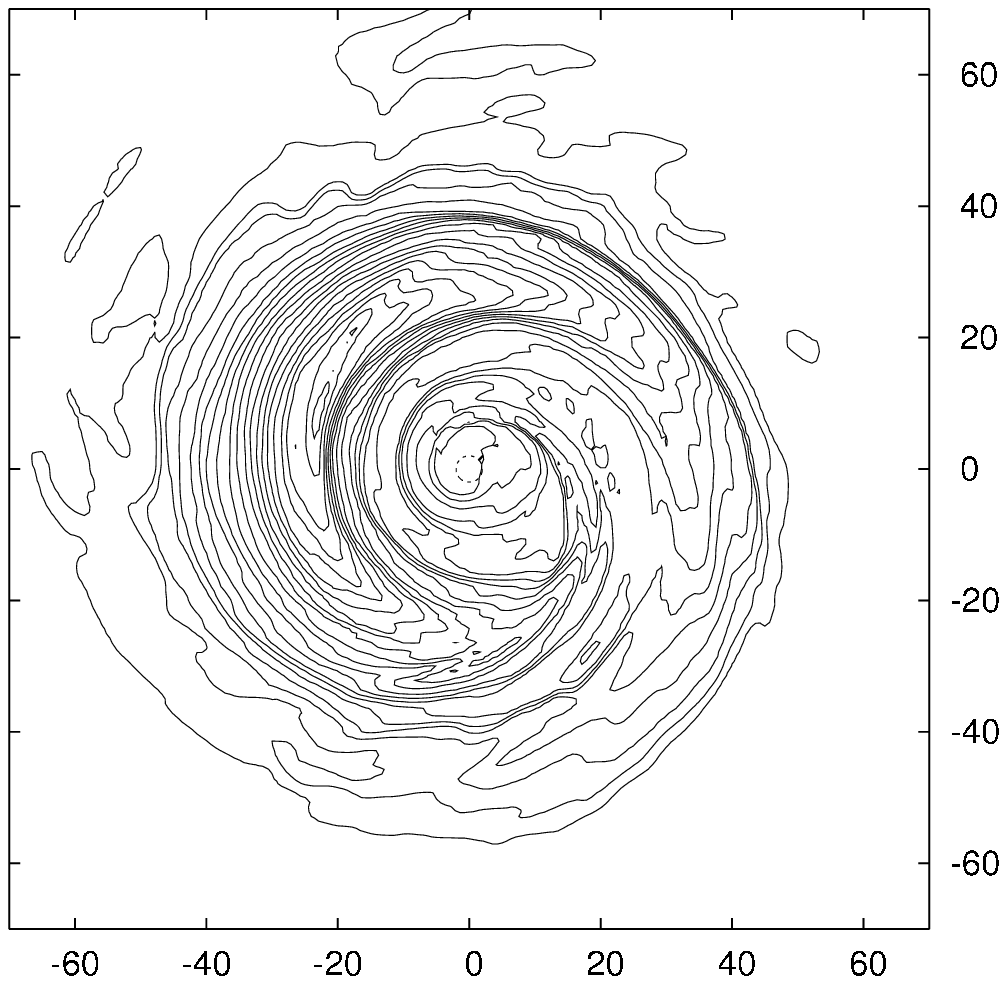}
\includegraphics[angle=0, width=0.49\columnwidth,clip]{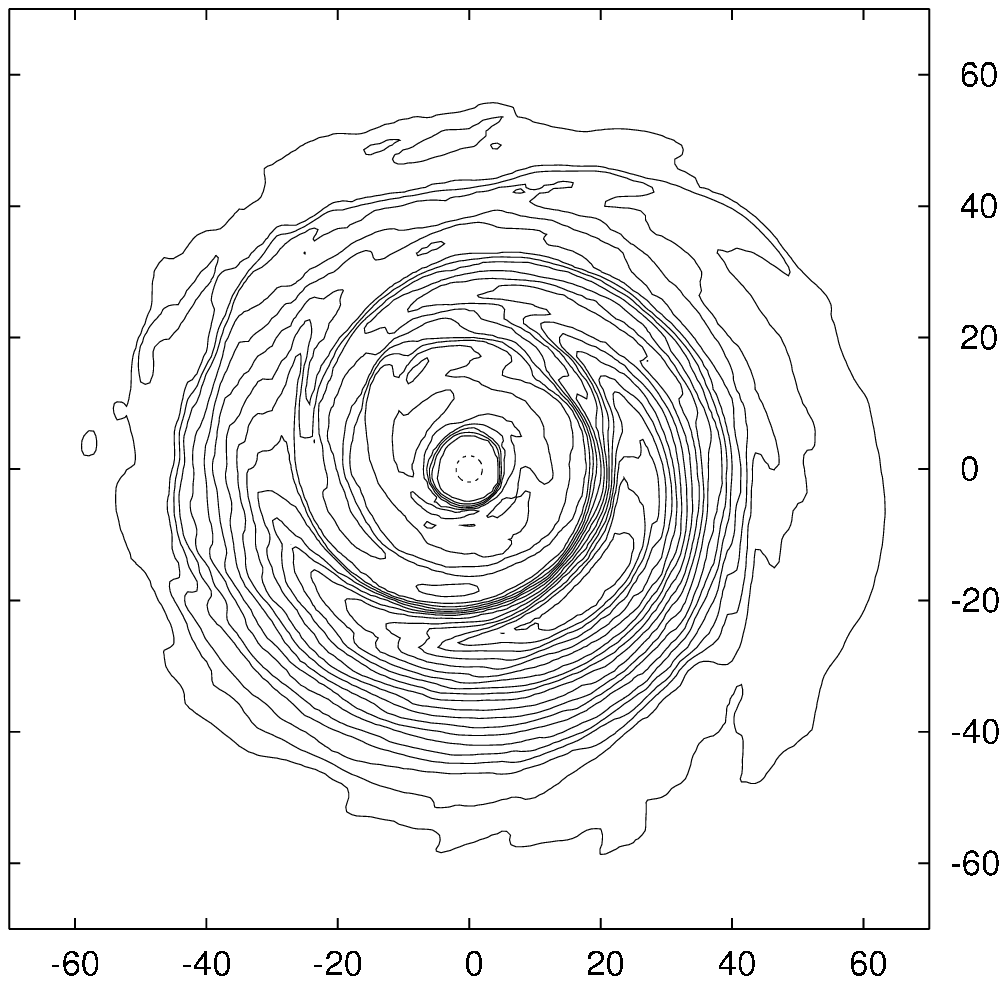}
\includegraphics[angle=0, width=0.49\columnwidth,clip]{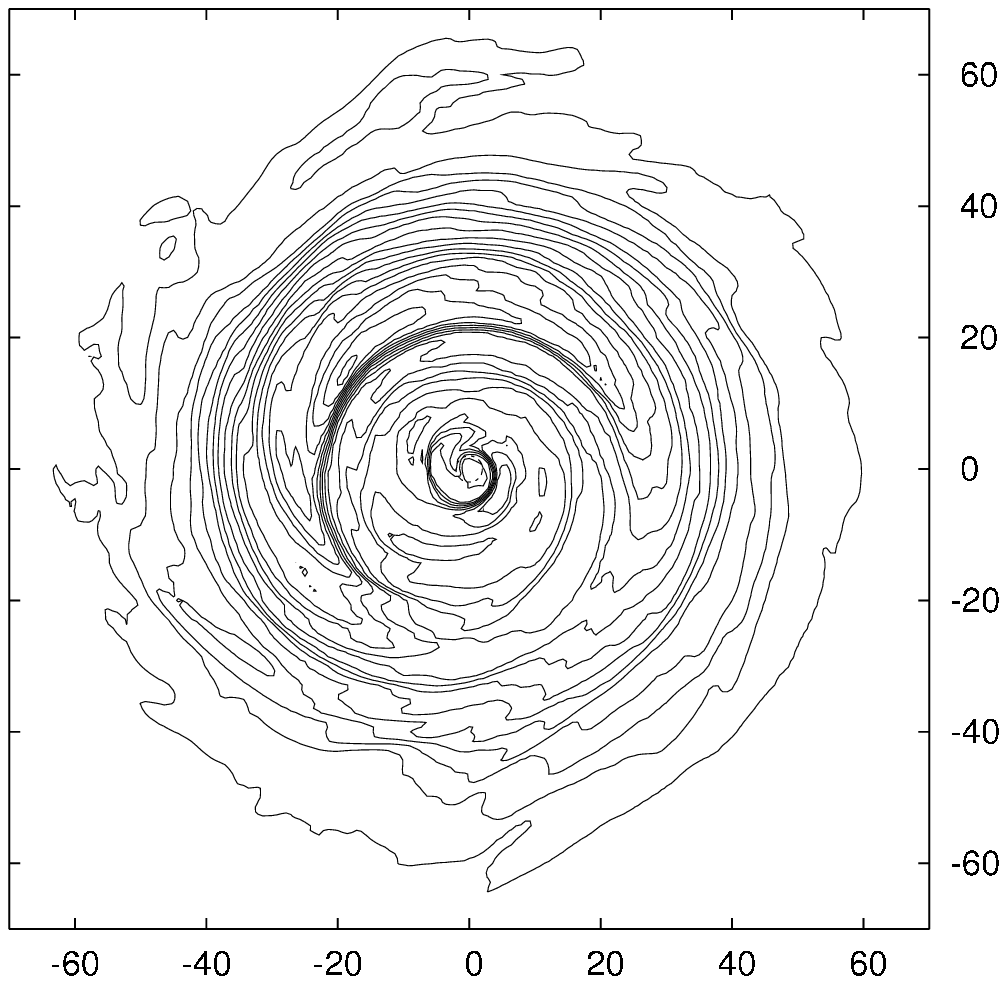}
\caption{Later stages of the simulation shown in figure~\ref{f:DVrhoF3}, in
  which the gas begins to accrete into the black hole. The
  panels show snaps from $t/M=6000$ (top left) to $7500$ (bottom right) at $\Delta
  t/M=500$ intervals. Notice that at $t/M=7000$ the ISCO is clearly noticeable.\label{f:accretion}}
 \end{center}
\end{figure}

Next we concentrate on the oblique recoil case.
For concreteness we adopt recoil velocity
values $v_{\rm kick}=300$, $1000$ and $3000$ km/s (we also consider 9000~km/s to
verify the appearance of the main feature and check the empirical law presented below).
We begin by examining
the case where the kick direction is on the orbital plane (i.e., orthogonal
to the angular momentum of the disk).
The simulations for the different cases proceed along qualitatively similar
phases, which are illustrated for the case of $v_{\rm kick}$=3000~km/s in
figure~\ref{f:DVrhoF3} for $\rho$ at $z=0$, and in figure~\ref{f:DVgradpF3} for
$|\nabla P|$ at $z=0$. The asymmetry introduced
by the kick's direction induces an accumulation of gas at one side of the disk, while
causing a significant decrease on the opposite side. As time progresses, shocks develop
and a complex dynamic arises, at late times $\simeq 6000M$, an accretion phase
is clearly noticeable for $v_{\rm kick} > $1000~km/s (see figure~\ref{f:accretion}).

\begin{figure}[ht!]
\begin{center}
\includegraphics[angle=0, width=0.49\columnwidth,clip]{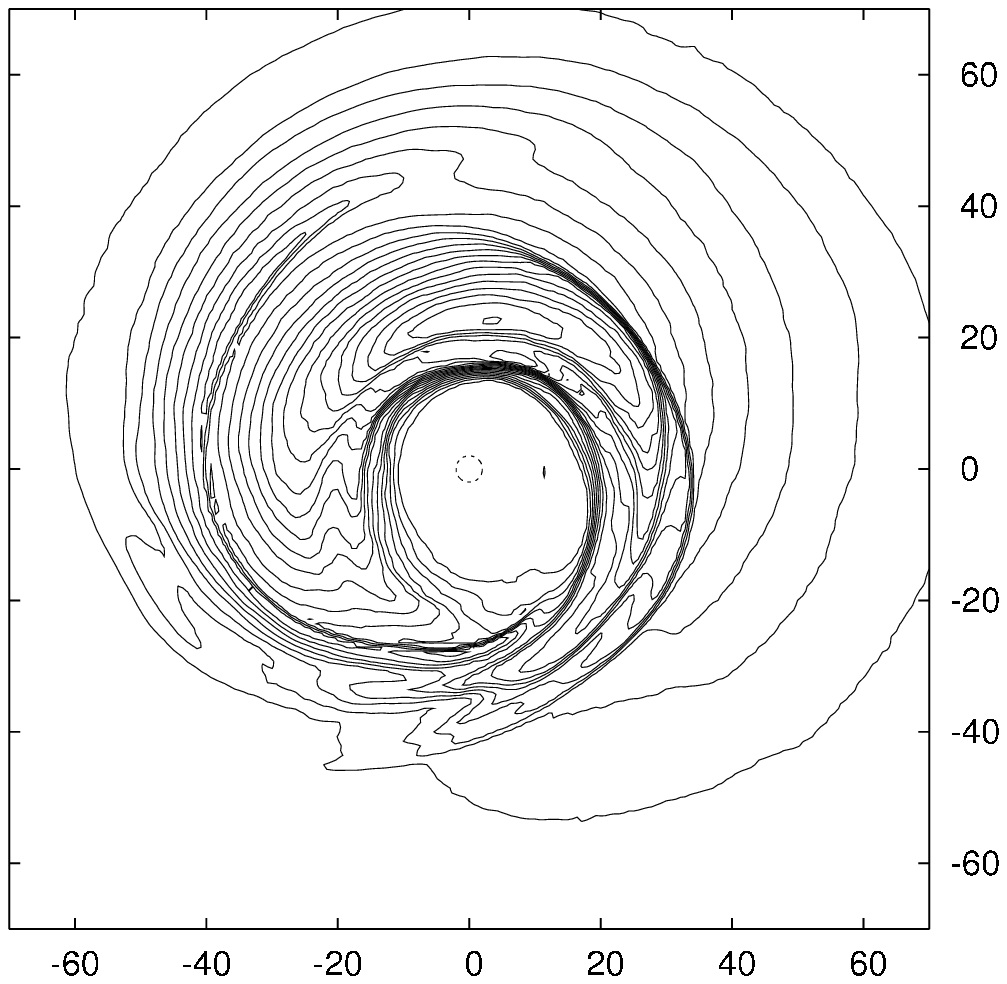}
\includegraphics[angle=0, width=0.49\columnwidth,clip]{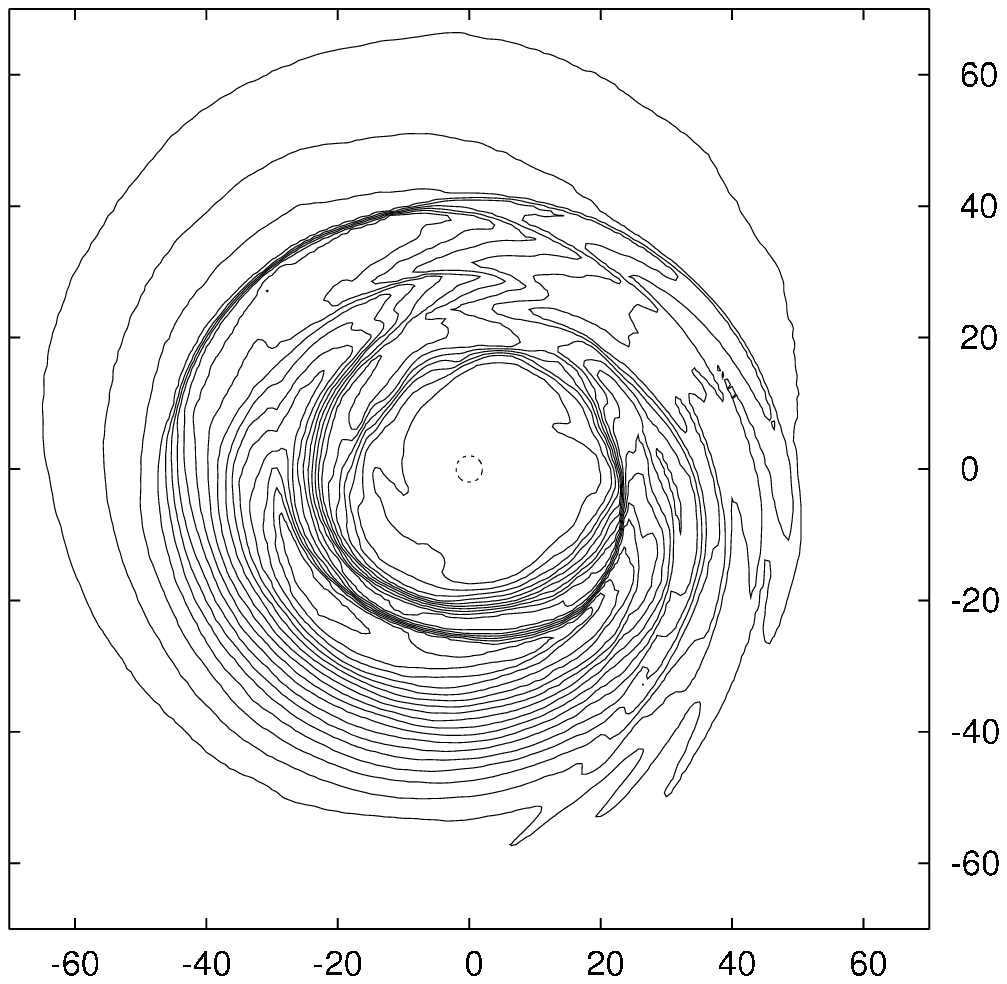}
\includegraphics[angle=0, width=0.49\columnwidth,clip]{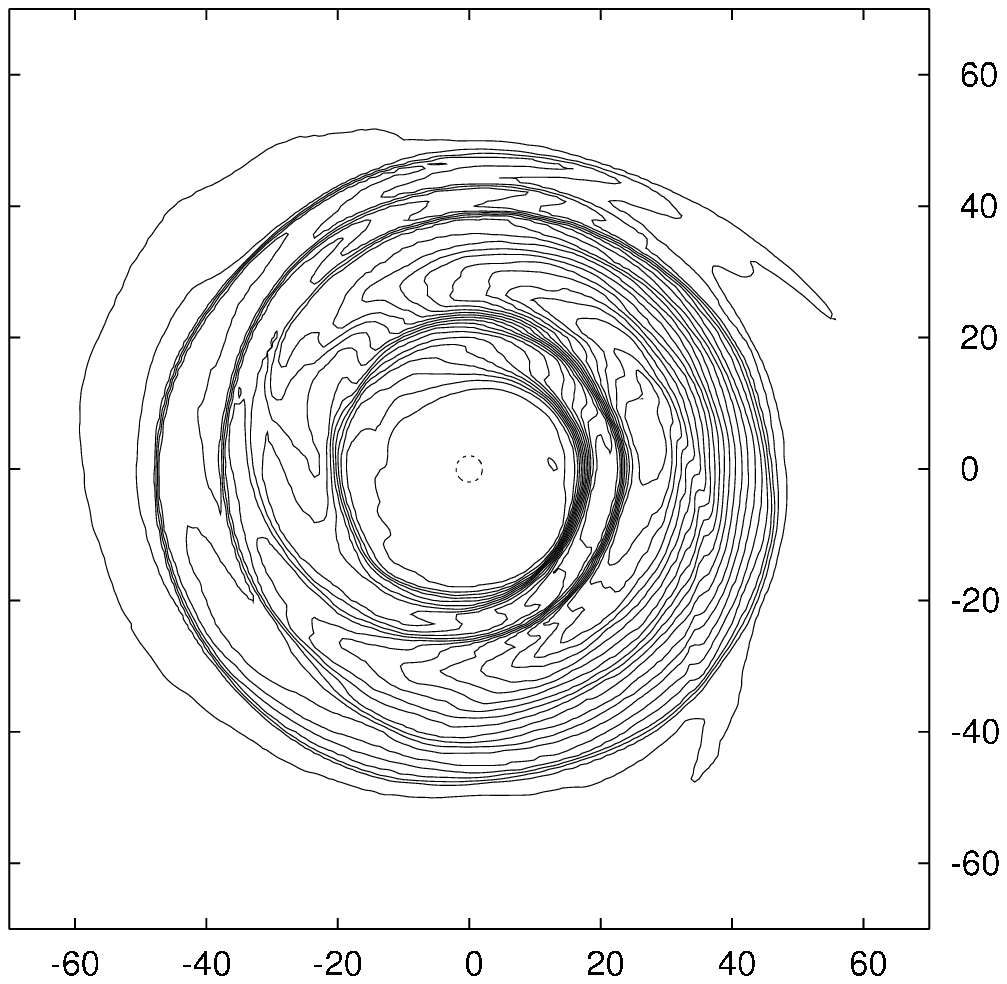}
\includegraphics[angle=0, width=0.49\columnwidth,clip]{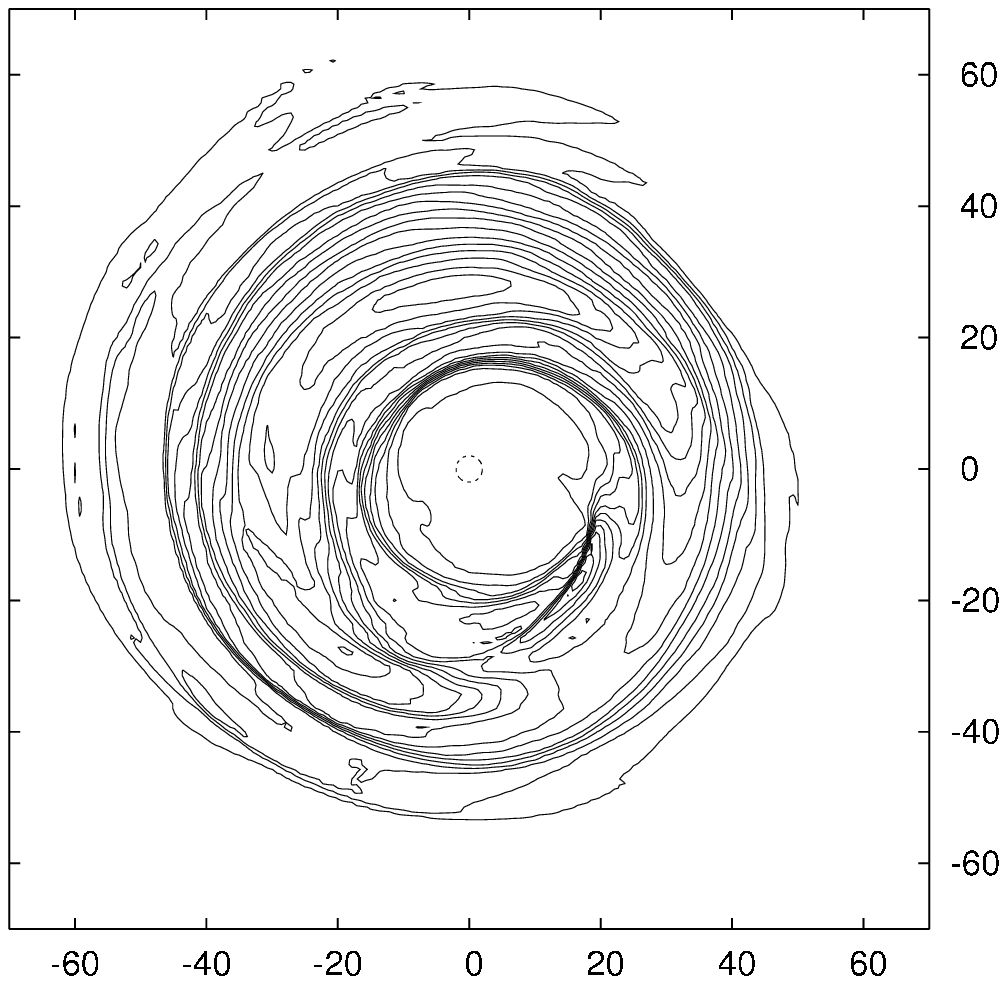}
\caption{Density $\rho$ at plane $z=0$ in the case of a disk kicked with a velocity
  3000~km/s at $\theta=30^\circ$. The panels show
  snaps from $t/M=2500$ (top left) to $4000$ (bottom right) at $\Delta t/M=500$
  intervals.\label{f:rho3000a30}}
 \end{center}
\end{figure}

\begin{figure}[ht!]
\begin{center}
\includegraphics[angle=0, width=0.9\columnwidth,clip]{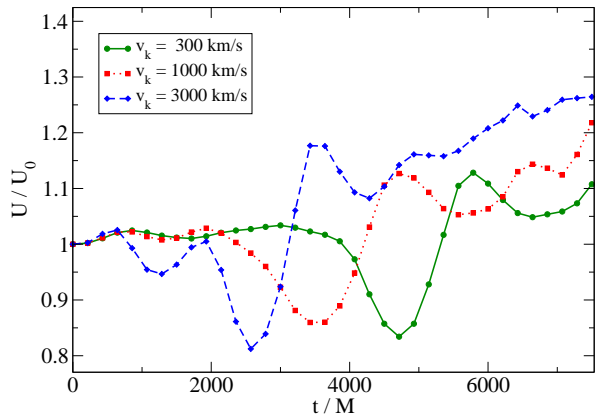}
\caption{Normalized internal energy. Kicks perpendicular to axis of rotation
  ($\theta=90^\circ$). \label{f:ie_magnitude}}
 \end{center}
\end{figure}

\begin{figure}[ht!]
\begin{center}
\includegraphics[angle=0, width=0.9\columnwidth,clip]{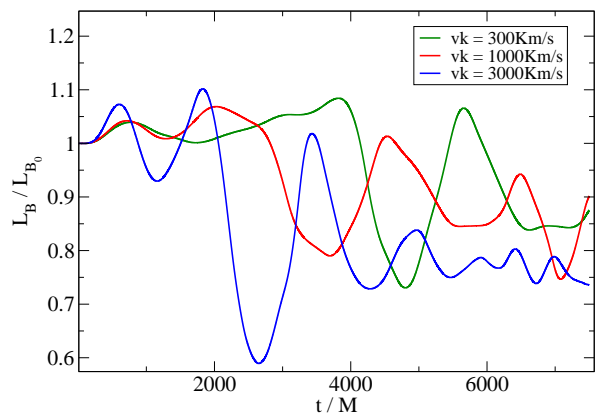}
\caption{Normalized bremsstrahlung luminosity. Kicks perpendicular to axis of rotation
  ($\theta=90^\circ$). \label{f:brem_magnitude}}
 \end{center}
\end{figure}

To analyze the impact of the disk dynamics and possible observable features,
we compute the internal energy (Fig.~\ref{f:ie_magnitude}) and bremsstrahlung
luminosity (Fig.~\ref{f:brem_magnitude}) for $v_{\rm kick}$=300, 1000, and 3000~km/s.
An initial relatively small bump is observed, which takes place at a time given by half the orbital
period of the maximum density region, which is consistent with the epicyclic picture.
From there on, a complex behavior is observed, though notably, irrespective of
the magnitude of the kick, the same qualitative features are observed  --especially
in Fig.~\ref{f:ie_magnitude}. Generally, we see that both the internal energy
and the bremsstrahlung luminosity dip and rebound but the internal energy ends
up higher, while the bremsstrahlung luminosity finishes lower. This can be
understood as follows: the kick energy is dissipated in shocks, increasing the
temperature and the pressure but the subsequent expansion reduces the density
below the initial values. Because the bremsstrahlung emissivity is
$\propto \rho^2 T^{1/2}$, the net effect is a reduction in emissivity despite
the increase in pressure. The relative changes in both internal energy and
bremsstrahlung luminosity are relatively modest, at a level of $\sim 20-40$\%
and occur on characteristic timescales on the order of
$1000 M=5000 M_6$ s, where $M_6$ is the mass of the black hole in $10^6 M_\odot$.

\begin{figure}[ht!]
\begin{center}
\includegraphics[angle=0, width=0.9\columnwidth,clip]{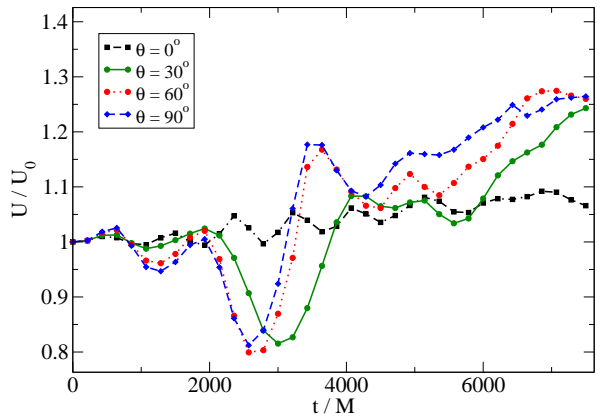}
\caption{Internal energy. Kicks at varying inclinations $\theta$ with respect to the
  axis of rotation. All cases with $v_{\rm kick}=3000$~km/s. \label{f:ie_angle}}
 \end{center}
\end{figure}

\begin{figure}[ht!]
\begin{center}
\includegraphics[angle=0, width=0.9\columnwidth,clip]{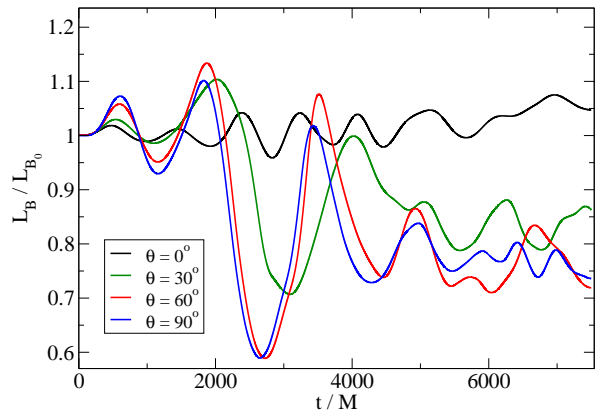}
\caption{Bremsstrahlung luminosity for kicks at varying inclinations $\theta$
  with respect to the axis of rotation. All cases with $v_{\rm kick}=3000$~km/s.
  \label{f:brem_angle}}
 \end{center}
\end{figure}

Second, we examine the dependence on kicks at different angles. Since the main
qualitative features of all kick cases considered are similar, we concentrate
in the case $v_{\rm kick}$=3000~km/s as this is the one that displays the
overall behavior within the shortest computational time. We compute the internal
energy and bremsstrahlung luminosity for kicks at $\theta=0, 30, 60,$ and
$90^\circ$, where the angle $\theta$ is measured with respect to the axis of rotation. 
Figure~\ref{f:rho3000a30} shows the density at plane $z=0$  in the case of a
kick at $30^\circ$. Figures \ref{f:ie_angle} and \ref{f:brem_angle} illustrate
the (normalized) internal energy and bremsstrahlung luminosity vs time for the
different angles considered. Recall that no significant shocks form when the
kick is along the axis of the disk. When the kick has a component along the
disk's plane however, the qualitative features observed in the internal energy
are similar for all cases. We note that the evolution we observe for a given 
$v_\perp = (3000~{\rm km/s}) \sin\theta$ is nicely bracketed by evolutions with pure
orbital plane kicks above and below $v_\perp$. Thus, $v_\perp$ is the most important
parameter determining the behavior of the kicked disk, apart from the small
oscillations also present when the kick is parallel to the axis of
rotation, and the likely small differences in the shape of the initial shock.

Another feature common to all the internal energy (or pressure) results (See Figs.
\ref{f:ie_magnitude} and \ref{f:ie_angle}) is a rapid swing from a dip to a
bump, followed by an oscillating growth at a moderate pace. While the
magnitude of the upward swing of the internal energy does not depend strongly
on $v_\perp$, the time at which it occurs does. The delay we observe decreases
as $v_\perp$ is increased.
If this delay were due to the time taken by a perturbation traveling at $v_\perp$
to cross some fixed distance, one would expect a dependence $\propto
v_\perp^{-1}$. Instead, we observe a logarithmic decrease.
Defining the delay as the time after the initial kick at which the internal
energy swings upward through the initial value, we find the following empirical dependence:
\begin{equation}
\frac{t_{\rm swing}}{M} = 5200 - 912 \ln{ \left( \frac{v_\perp}{300~{\rm km/s}}
                                          \right) }
\label{tswing}
\end{equation}
Note that this formula applied na\"ively ``predicts" an infinite delay for a kick along the axis of rotation.

As is well known, constant specific angular momentum tori are prone to a
nonaxisymmetric corotation instability \cite{papa_pringle,1986MNRAS.221..339G} whose nonlinear development has been explored
numerically in the pseudo-Newtonian approximation \cite{Hawley:2000} and in
GRMHD \cite{DeVilliers:2002ab}.
The final outcome depends on
the aspect ratio of the torus, the nature and strength of any large-scale
magnetic fields present, the presence of accretion \cite{blaes},
and this remains to be fully investigated in GRMHD context.
Therefore, any substantial perpendicular component of the kick is likely to excite at
some level the $m=1$ nonaxisymmetric mode, which is expected to grow at a rate
$\omega\approx 0.2 \Omega_{\rm m}$, where $\Omega_{\rm m}= 2\pi/P_{\rm m}$ is
the Keplerian angular frequency at the pressure maximum. For the parameters of
the torus of our simulations $P_{\rm m}=1220 M$, and
$\Omega_{\rm m}= 0.00515$. The behavior described above is suggestive: if one
assumes that the initial pressure perturbation is $\delta P_0\propto
v_\perp^2$, which is reasonable for shocks and on dimensional grounds, and one
sets $\delta P = \delta P_0 \exp{\omega t}$, then the time required for the
perturbation to attain a given fiducial level would follow an equation of the
form (\ref{tswing}), with $t=t_{\rm ref}-(2/\omega)\ln{(v_\perp/v_{\rm
    ref})}$, where $t_{\rm ref}$ and $v_{\rm ref}$ are some arbitrary reference values.
Analyzing the results we obtained indicates that $\omega=0.43\Omega_{\rm
    m}$, which is on the order of the expected frequency but significantly
  higher. Thus, we suggest tentatively that the swing we see in both the
  internal energy and bremsstrahlung plots in all cases where there is a
  nonzero $v_\perp$ is a common transient response to the kick that may be
  observable in principle, and that the subsequent growth may be due to the
  growth of the instability and/or the rise to the expected level of
  dissipation of the input kinetic energy. At late times for the higher kicks
  our simulations display an accretion phase and so this possible saturation can not
  be explored, though a suggestive behavior consistent with this saturation is displayed
by the largest kick considered.

We note that a similar swing in the bremsstrahlung luminosity was
 observed in the (axisymmetry preserving) simulations by O'Neill et al \cite{massloss} using thin disks, which
 are not prone to the Papaloizou and Pringle instability. In the
 near future, to further elucidate the relative importance of the transient
 response and the instability,
we are planning an investigation of the effects of kicks in tori with flatter rotation laws
$\Omega\propto r^{-q}$ since the aforementioned instability does not occur
if $q<\sqrt{3}$ as well as examining magnetized tori.

\subsection{Dependence on black hole spin}

\begin{figure}[ht!]
\begin{center}
\includegraphics[angle=0, width=0.9\columnwidth,clip]{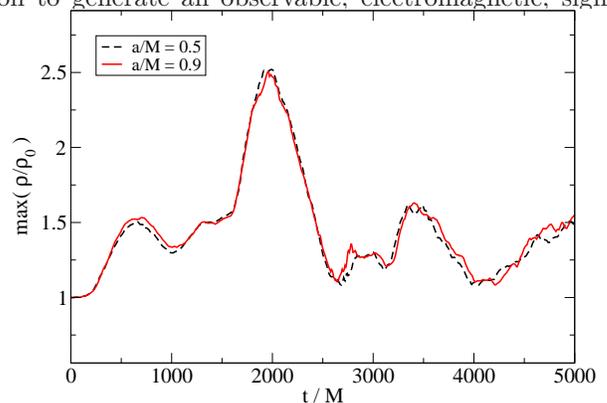}
\caption{Comparison of the maximum of density for black hole spins
  $a/M=0.5$ (dashed line) and $0.9$ (continuous line) for a kick of 3000~km/s
  perpendicular to the disk's axis.  \label{f:spin_comp}}
 \end{center}
\end{figure}

Finally, we investigate possible differences between cases with different black hole
spins by performing a simulation with spin $a/M = 0.9$
in addition to the value $a/M = 0.5$ used in the rest of the simulations. For
this test we chose the setting with kick velocity of 3000~km/s perpendicular to
the disk's axis. Notice that although all other parameters coincide in these
simulations, the stationary disk solutions used to construct the initial data
are slightly different since they depend on $a$. Still, we see no significant
differences, as is illustrated
in figure~\ref{f:spin_comp}, where we show a comparison between the maxima of
density, normalized by dividing by its value at $t=0$, which is slightly
different in each case.

%
%
\section{Conclusion}
In the current work we have studied the possibility that binary black hole
mergers, within a circumbinary disk, give rise to scenarios likely to emit
electromagnetic radiation. We have studied both the impact of mass loss in
the system and recoil velocities. While both induce deformations of the disk,
it is the case of a recoiling black hole, when the recoil's direction has
a component along the disk plane, that appears as the most promising option
to generate an observable, electromagnetic, signature.
This is so not just because the effect
is larger, but also the variability induced is significantly more pronounced
than that observed in the case of mass loss or kick along the disk's angular momentum.
Furthermore, we find that the magnitude of the kick is not very important as
long as it is less than the smallest orbital speed of the fluid. While the kick magnitude impacts the time at which
the strongest variation in internal energy or bremsstrahlung appears, the intensity
and time scale of the variation and behavior afterwards is not. Since supermassive
binary black hole mergers will generically
give rise to recoils (simply by having a mass ratio different from unity) which
in turn ensures a kick component orthogonal to the final black hole spin, effects
like those observed here indicate a possible a common behavior for the majority of scenarios.

Our studies also indicate that the final black hole spin has no strong effect
on the main features of the solution. 
However, if an accretion phase takes place, the
location of the innermost stable circular orbit will naturally play a key role.

%
%
\noindent{\bf{\em Acknowledgments:}}
We would like to thank
P. Chang, B. Kocsis, J. McKinney, C. Miller, S. Phinney and J. Tohline for stimulating
discussions.
This work was supported by the NSF grants
PHY-0803629, PHY-0653375 and NASA ATP grant NNX07AG84G to LSU,  PHY-0803615 and CCF-0832966
to BYU, and PHY-0803624 and CCF-0833090 to LIU.
Computations were done at BYU (Marylou4), the Louisiana
Optical Network Initiative (LONI), LSU, and
TeraGrid.
LL acknowledges the Aspen Center for Physics for hospitality where
this work was started.

%
%

%
%
\end{document}